\documentclass[lettersize,journal]{IEEEtran}
\usepackage{amsmath,amsfonts}
\usepackage{algorithmic}
\usepackage{array}
\usepackage[caption=false,font=normalsize,labelfont=sf,textfont=sf]{subfig}
\usepackage{textcomp}
\usepackage{stfloats}
\usepackage{url}
\usepackage{verbatim}
\usepackage{graphicx}
\def\BibTeX{{\rm B\kern-.05em{\sc i\kern-.025em b}\kern-.08em
    T\kern-.1667em\lower.7ex\hbox{E}\kern-.125emX}}
\usepackage{balance}
\usepackage{cite}

\usepackage{algorithm,algorithmic}
\usepackage{float}
\usepackage[caption = false]{subfig}

\usepackage{bm}
\usepackage{xcolor}
\usepackage{color}
\definecolor{UPC}{RGB}{0,122,191}
 \usepackage[american]{circuitikzgit}
\ctikzset{bipoles/length=1cm}
 \usepackage{tikz,tkz-graph}
 \usetikzlibrary{shapes,arrows}
 \usetikzlibrary{calc}
 \usetikzlibrary{fit}
 
\pgfdeclarelayer{background}
\pgfsetlayers{background,main}

\begin{document}
\title{Implementation of complex-valued sliding mode controllers in three-phase power converters}
\author{Arnau Dòria-Cerezo, Pau Boira, Víctor Repecho and Domingo Biel
\thanks{This work was partly supported by the Government of Spain through the Agencia Estatal de Investigación under Project PID2021-122821NB-I00, the Generalitat de Catalunya through Project 2021 SGR 00376.

Arnau Dòria-Cerezo is with the Department of Electrical Engineering and the Institute of Industrial and Control Engineering, Universitat Politècnica de Catalunya, 08028 Barcelona, Spain (e-mail: arnau.doria@upc.edu).

Pau Boira is with the Institute of Industrial and Control Engineering, Universitat Politècnica de Catalunya, 08028 Barcelona, Spain (e-mail: pau.boira@upc.edu).

Víctor Repecho is with the Department of Automatic Control and the Institute of Industrial and Control Engineering, Universitat Politècnica de Catalunya, 08028 Barcelona, Spain (e-mail: victor.repecho.del@upc.edu).

Domingo Biel is with the Department of Electronics and the Institute of Industrial and Control Engineering, Universitat Politècnica de Catalunya, 08028 Barcelona, Spain (e-mail: domingo.biel@upc.edu).
}
}

\maketitle

\begin{abstract}
This paper presents two methods for implementing complex-valued sliding mode controllers in three-phase power converters. The paper includes the description of the algorithms and a detailed analysis of the proposed implementations. The methods, that are easy to code and have a low computational burden, retain the sliding mode properties of robustness and fast response and do not require any additional processing often used to decouple the dynamics of the three-phase system. The performance of the methods is compared in numerical simulations, and the algorithms are experimentally tested in a microcontroller using a Hardware-in-the-Loop platform.
\end{abstract}


\section{Introduction} \label{intro}

\IEEEPARstart{T}{he} implementation of controllers in power electronic converters usually requires a modulation stage that translates the continuous control actions calculated by the control algorithms into the switching policy that defines the switch combinations. Pulse-Width Modulation (PWM), and its variations, is the most used approach in single-phase systems and Space-Vector Modulation (SVM) is a common alternative applied in three-phase power converters \cite{HL2003}.

These implementation strategies are well known and compare, at some point in the control-to-time conversion, the control signal with a carrier waveform to get the on/off time intervals of the power switches. Due to their inherent operation, PWM or SVM can be used with controllers designed using linear techniques or non-linear controllers that provide a continuous signal at their output. Alternatively, discontinuous nonlinear techniques such as bang-bang (or hysteresis) control or Sliding Mode Control (SMC) \cite{UGS1999}, which do not require a modulation stage, have been successfully applied to single-phase power electronics systems offering excellent robustness performances with excellent dynamics. The reader can find some examples in \cite{UGS1999}. The use of sliding modes for complex-valued systems has recently been proposed for control three-phase electrical systems. See examples in the control of induction motors \cite{DORB2020,DHB2022,SMTSJB2023} or in three phase-locked loop algorithms \cite{DRB2021}. Compared to traditional \emph{real-valued} SMC, the advantages of the complex-valued SMC (cSMC) approach are a faster response, the compactness in the solution, and a different perspective resulting in alternative solutions often difficult to find using \emph{real-valued} approaches \cite{DOBF2021}. The main motivation for using the complex-valued description is the compactness of the analysis because multi-input multi-output (MIMO) systems can be reduced to single-input, single-output (SISO) relationships \cite{Harnefors2007}.

Regarding SMC implementation, there are many proposals in the specialised literature for single-phase power converters \cite{HDM2013,HCC2009,MMT1997,RBFG2003,QLTH2018,AAM2012,HYLHC2013,ZJSGT2018,VMMS2018,PR2017,SVKMA2016,TC2002,RBO2021}.
Unfortunately, these solutions cannot be directly applied to three-phase systems. Three-phase power converter dynamics can be described in different frameworks, from the \emph{abc} to the \emph{dq} description applying the Clarke/Park transformations. The SMC design in the \emph{dq}-framework is simplest because the control objective is simplified to an output regulation problem. However, to be applied to the power converter, the control action in the \emph{dq} coordinates must be transformed to three-phase variables in \emph{abc}, which complicates the control implementation for SMCs with discontinuous output signals, see \cite{UGS1999}. Therefore, the \emph{dq}-framework is used by SMC with continuous outputs, such as Super Twisting SMC, \cite{LSGHG2020,OBBBS2021} among others, where SVM can be applied eventually.
Regarding the SMC designed in the \emph{abc}-framework, the design procedure is not straightforward because the state equations of the three-phase system are not decoupled \cite{UGS1999}. Several approaches can be found in the literature that tackle this problem. Among them can be cited the work in \cite{GGMCM2016}, in which a Kalman filter is designed to decouple the equations of a three-phase rectifier, the article \cite{AOKS2019} in which the sliding mode control law is obtained in a three-phase grid-tied LCL-interfaced inverter without applying any decoupling procedure and which ensures the sliding regime under certain assumptions, the paper \cite{RBA2018} in where the SMC is designed using a decoupling matrix for driving a permanent magnet synchronous machine, or the work proposed in \cite{MGGCM2018} which defines six regions and the switch positions are selected using a nonlinear matrix equation. The use of sectors has also been proposed to implement the cSMC in \cite{DHB2022} since this simple solution naturally arises when the complex-valued signals are represented in the complex plane.   


This paper proposes two methods to implement the controllers designed for three-phase inverters using cSMC in a microcontroller. The algorithms use the sampling time of the digital implementation and average the desired discontinuous control action between available switching combinations. The procedure looks similar to the SVM technique because of the use of sectors but is inherently different because there are no modulation purposes. With respect to the mentioned strategies to implement sliding mode controllers, the proposed methods are easier to implement and admit the use of zero vectors, which offers new possibilities for controlling power converters.

The paper is organised as follows. Section \ref{sec_cSMCrevisited} revisits the complex-valued sliding mode technique. The relationship between the complex-valued signals and the available switch combinations in a three-phase inverter is discussed in Section \ref{sec_SMCCSS}.  Then, Section \ref{sec_SampledMethods} introduces the sampled-time implementation methods proposed in the paper. In Section \ref{sec_example}, a Voltage Source Inverter (VSI) is used as an example to test and compare the proposed strategies. Finally, the conclusions are stated in Section \ref{sec_concl}.

\subsection{Notation}

$j=\sqrt{-1}$ is used, instead of $i$, to avoid confusion with electrical currents; $\mathbb{C}^n$ denotes the complex $n$th-dimensional space;  $\bm{z}\in\mathbb{C}$ denotes a complex-valued variable; for a complex value, $\bm{z}$, $\mathrm{Re}(\bm{z})$ and $\mathrm{Im}(\bm{z})$ denote the real and imaginary parts, respectively, $\bar{\bm{z}}$, $|\bm{z}|=Z$, and $\delta_z$ denote the conjugate, magnitude, and argument (angle), respectively, of $\bm{z}\in\mathbb{C}$.

\section{Complex-valued Sliding Mode Control}\label{sec_cSMCrevisited}

This section briefly introduces the sliding mode control technique for complex-valued systems presented in \cite{DOBF2021}. Consider a single input complex-valued system described by
\begin{equation}
\dot{\bm{z}}=f(\bm{z})+g(\bm{z})\bm{u},\label{eq_sys}
\end{equation}
where $\bm{z}\in\mathbb{C}^n$ is the state vector, and $f,g$ are complex valued functions, and $\bm{u}\in\mathbb{C}$ is the control input.

The sliding mode control technique for a complex-valued system consists of defining a complex switching function, $\bm{\sigma}(\bm{z},t)$, and forcing the sliding motion, i.e., driving and maintaining the system \eqref{eq_sys} on the switching manifold $\bm{\sigma}=0$. First-order sliding mode control requires that $\bm{\sigma}(\bm{z})$ is a relative degree one with respect to the control input.

Once the switching function $\bm{\sigma}(\bm{z},t)$ is selected according to the control objectives, the design of complex sliding mode controllers can be divided into three steps:

\begin{enumerate}
\item \emph{Equivalent control.} The equivalent control, $\bm{u}_{eq}$, is defined so that $\bm{\sigma}=0$ and $\dot{\bm{\sigma}}=0$, and provides the required control action that ensures remaining on the switching manifold. Differentiating $\bm{\sigma}(\bm{z},t)$, with respect to time, and using \eqref{eq_sys},
\begin{equation}
\dot{\bm{\sigma}}=\frac{\partial \bm{\sigma}}{\partial \bm{z}}\left(f(\bm{z})+g(\bm{z})\bm{u}\right)+\frac{\partial\bm{\sigma}}{\partial t}
\end{equation}
and, the equivalent control, with $\dot{\bm{\sigma}}=0$, results as
\begin{equation}
\bm{u}_{eq}=-\left(\frac{\partial \bm{\sigma}}{\partial \bm{z}}g(\bm{z})\right)^{-1}\left(\frac{\partial \bm{\sigma}}{\partial \bm{z}}f(\bm{z})+\frac{\partial\bm{\sigma}}{\partial t}\right),\label{eq_ueq}
\end{equation}
where $\left|\frac{\partial \bm{\sigma}}{\partial \bm{z}}g(\bm{z})\right|\neq0$, known as the transversality condition, must be fulfilled.

\item \emph{Ideal sliding dynamics.} The internal stability on the sliding surface must be guaranteed. The remaining dynamics once the sliding motion is reached, known as the ideal sliding dynamics, are defined by
\begin{equation}
\dot{\bm{z}}=f(\bm{z})+g(\bm{z})\bm{u}_{eq},
\end{equation}
subject to $\bm{\sigma}(\bm{z},t)$.

\item \emph{Switching control law.} The proposed switched control action is 
\begin{equation}
\bm{u}=-\bm{\kappa}\frac{\bm{\sigma}}{|\bm{\sigma}|},\label{eq_u}
\end{equation}
where $\bm{\kappa}\in\mathbb{C}$ fulfills 
\begin{equation}
|\bm{\kappa}|\cos(\delta_{\sigma_g}+\delta_\kappa)>|\bm{u}_{eq}|.\label{eq_cond}
\end{equation}
with $\delta_{\sigma_g}$ and $\delta_\kappa$, the arguments of $\frac{\partial \bm{\sigma}}{\partial \bm{z}}g(\bm{z})$ and $\bm{\kappa}$, respectively. With this control law, the sliding motion is guaranteed with the Lyapunov function candidate
\begin{equation}
V=\frac{1}{2}\bar{\bm{\sigma}}\bm{\sigma}.\label{eq_V}
\end{equation}
Note that, the time derivative of \eqref{eq_V} can be written as
\begin{equation}
\dot{V}=\mathrm{Re}\left(\bar{\bm{\sigma}}\dot{\bm{\sigma}}\right),
\end{equation}
that, using \eqref{eq_sys}, \eqref{eq_u}, \eqref{eq_ueq}, results in
\begin{align}
\dot{V}=&-|\bm{\sigma}|\left|\frac{\partial \bm{\sigma}}{\partial \bm{z}}g(\bm{z})\right|\left(|\bm{\kappa}|\cos(\delta_{\sigma_g}+\delta_\kappa)\right.\nonumber\\
&\left.+|\bm{u}_{eq}|\cos(\delta_{\sigma_g}-\delta_\sigma+\delta_{eq})\right),
\end{align}
and, condition \eqref{eq_cond} ensures $\dot{V}<0$, and the convergence to $\bm{\sigma}=0$.

\end{enumerate}

Control action in \eqref{eq_u} results in the set of points defined by a circumference of radius $|\bm{\kappa}|$ in the complex plane; see an example in the blue circumference in Figure \ref{fig_2level}.

\section{Complex-valued SMC in three-phase power converters} \label{sec_SMCCSS}

\subsection{Transformation from three-phase electrical variables to complex variables}

The transformation from three-phase to complex-valued variables is equivalent to the well-known $abc$ to $\alpha\beta$ transformation; see more details in \cite{Harnefors2007}. The complex transformation of any periodic three-phase signal $x_{abc}(t)\in\mathbb{R}^3$ is defined by
\begin{equation}
\begin{pmatrix}\bm{x}(t)\\\bar{\bm{x}}(t)\\x_0(t)\end{pmatrix}=\bm{T}\begin{pmatrix}x_a(t)\\x_b(t)\\x_c(t)\end{pmatrix},\label{eq_ComplexTransf}
\end{equation}
where $x_k(t)$, with $k=a,b,c$, are the (real-valued) three-phase variables, $\bm{x}(t)$ is the complex-valued variable, $x_0$ is the (real-valued) homopolar component and
\begin{equation}
\bm{T}=c\begin{pmatrix}1&\bm{\alpha}&\bar{\bm{\alpha}}\\1&\bar{\bm{\alpha}}&\bm{\alpha}\\1&1&1\end{pmatrix}\in\mathbb{C}^{3\times 3}
\end{equation}
where $\bm{\alpha}=e^{j\frac{2\pi}{3}}$ and $c$ could take different values. The most used ones are $c=\frac{2}{3}$ for amplitude preserving, or $c=\frac{1}{\sqrt{3}}$ that preserves the instantaneous electrical power defined as $p(t)=\mathrm{Re}(\bm{v}\bar{\bm{i}})$, where $\bm{v},\bm{i}$ are the complex-valued voltage and current, respectively.

Notice that the two first components of the complex vector in \eqref{eq_ComplexTransf} are $\bm{x}(t)$ and $\bar{\bm{x}}(t)$, which are complex conjugate, and thus, redundant. In some cases, such as balanced conditions or the currents in three-wire systems, the homopolar component is $x_0(t)=0$.

Sometimes, the complex variables can be oriented to a reference frame (also known as the $dq$-model). Then
\begin{equation}
\bm{x}_{dq}(t)=e^{-j\theta}\bm{x}(t),\label{eq_albe2dq}
\end{equation}
where $\bm{x}_{dq}(t)\in\mathbb{C}$ and $\theta(t)$ corresponds to the angle of the desired reference.

\subsection{Available control actions in three-phase power converters}

\begin{figure}
\centering
\scalebox{0.75}{\begin{circuitikz}

\draw (0,0) node[anchor=east] {$V_{dc}$} to [short,o-*]++(1.5,0) coordinate (aux1)
	to [short,-*]++(1.5,0) coordinate (aux2)
	to [short]++(1.5,0) coordinate (aux3);

\draw (aux1) to [short]++(0,-0.25) node[nigbt,bodydiode, anchor=C](Sa){} (Sa.B) node[anchor=south] {$S_a^+$}
(Sa.E) to [short,-*]++(0,-0.25) coordinate(outA) to [short]++(0,-1.75)
to [short]++(0,-0.25) node[nigbt,bodydiode, anchor=C](SaN){} (SaN.B) node[anchor=south] {$S_a^-$}
(SaN.E) to [short,-*]++(0,-0.25);

\draw (aux2) to [short]++(0,-0.25) node[nigbt,bodydiode, anchor=C](Sb){} (Sb.B) node[anchor=south] {$S_b^+$}
(Sb.E) to [short,-*]++(0,-1) coordinate(outB) to [short]++(0,-1.0)
to [short]++(0,-0.25) node[nigbt,bodydiode, anchor=C](SbN){} (SbN.B) node[anchor=south] {$S_b^-$}
(SbN.E) to [short,-*]++(0,-0.25);

\draw (aux3) to [short]++(0,-0.25) node[nigbt,bodydiode, anchor=C](Sc){} (Sc.B) node[anchor=south] {$S_c^+$}
(Sc.E) to [short,-*]++(0,-1.75) coordinate(outC) to [short]++(0,-0.25)
to [short]++(0,-0.25) node[nigbt,bodydiode, anchor=C](ScN){} (ScN.B) node[anchor=south] {$S_c^-$}
(ScN.E) to [short]++(0,-0.25)
to [short,-o]++(-4.5,0) node[anchor=east] {$-V_{dc}$};

\draw (outA) to [short,-o]++(4.5,0) node[anchor=west] {$v_a'$}
	(outB) to [short,-o]++(3,0) node[anchor=west] {$v_b'$}
	(outC) to [short,-o]++(1.5,0) node[anchor=west] {$v_c'$};

\end{circuitikz} }
\caption{Three-phase bridge inverter.}\label{fig_3PhaseInverter}
\end{figure}
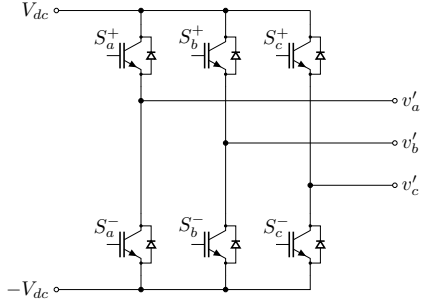

As mentioned in Section \ref{sec_cSMCrevisited}, the cSMC strategy results in switching between the values on a circle with radius $|\bm{\kappa}|$. However, when using a three-phase bridge inverter, see Figure \ref{fig_3PhaseInverter}, only a set of values defined by the on/off combination of switches is available.

Let us assume a VSI with three inductors in series on the AC side. Then, the switching policy must ensure the following:
\begin{itemize}
\item For the same $k$-phase with $k=a,b,c$, switches $S_k^+$ and $S_k^-$, can not be closed simultaneously to prevent short-circuiting the DC side of the power converter.
\item It is required to close one switch of each $k$-phase to ensure continuity of the current flow in the AC side of the converter.
\end{itemize}
The two previous conditions reduce the admissible switch positions to the eight combinations shown in Table \ref{table_Admissible1}. The six combinations named $V_n$ with $n = 1,\ldots, 6$ are called the \emph{active vectors}, and the two combinations $V_0^+$ and $V_0^-$ are called \emph{zero vectors}.

\begin{table}[htbp]
\centering
\begin{tabular}{||c||c|c|c|c|c|c|c|c||}
	\hline 
~ & $V_1$ & $V_2$  & $V_3$  & $V_4$ & $V_5$ & $V_6$ & $V_0^+$ & $V_0^-$\\ 
	\hline 
$S_a^+$ & $1$ & $1$ & $0$ & $0$ & $0$ & $1$ & $1$ & $0$\\ 
	\hline 
$S_a^-$ & $0$ & $0$ & $1$ & $1$ & $1$ & $0$ & $0$ & $1$\\ 
	\hline 
$S_b^+$ & $0$ & $1$ & $1$ & $1$ & $0$ & $0$ & $1$ & $0$\\ 
	\hline 
$S_b^-$ & $1$ & $0$ & $0$ & $0$ & $1$ & $1$ & $0$ & $1$\\
	\hline 
$S_c^+$ & $0$ & $0$ & $0$ & $1$ & $1$ & $1$ & $1$ & $0$\\
	\hline 
$S_c^-$ & $1$ & $1$ & $1$ & $0$ & $0$ & $0$ & $0$ & $1$\\
	\hline
\end{tabular} \vspace{0.1cm}
\caption{Admissible switch combinations.}
\label{table_Admissible1}
\end{table}

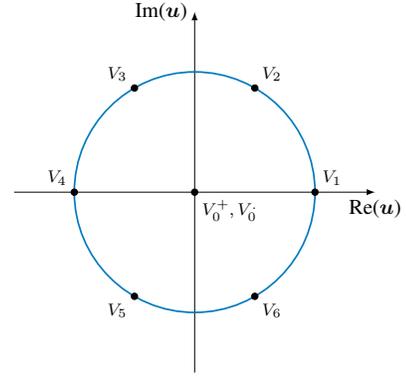
\begin{figure}
\centering
\scalebox{0.8}{
\begin{circuitikz}
\draw[-latex] (-3,0) -- (3,0) node[anchor=north]  {Re($\bm{u}$)};
\draw[-latex] (0,-3) -- (0,3) node[anchor=east]  {Im($\bm{u}$)};
\draw[thick,UPC] (0,0) circle (2);
\filldraw[black] 
(0,0) circle (1.5pt) node[anchor=north west] {\footnotesize $V_0^+,V_0^.$}
(0:2) circle (1.5pt) node[anchor=south west] {\footnotesize $V_1$}
(60:2) circle (1.5pt) node[anchor=south west] {\footnotesize $V_2$}
(120:2) circle (1.5pt) node[anchor=south east] {\footnotesize $V_3$}
(180:2) circle (1.5pt) node[anchor=south east] {\footnotesize $V_4$}
(-120:2) circle (1.5pt) node[anchor=north east] {\footnotesize $V_5$}
(-60:2) circle (1.5pt) node[anchor=north west] {\footnotesize $V_6$}
;
\end{circuitikz}
}
\caption{Complex control values from \eqref{eq_u} with a radius $|\bm{\kappa}|$ (blue line). Black dots are the mapped values of the switching combinations of a three-phase bridge converter in \eqref{eq_uabc} using transformation \eqref{eq_ComplexTransf}, with a distance to the origin of $2c$.}\label{fig_2level}
\end{figure}

Defining the control variable as $u_k = S_k^+ - S_k^-$ for $k = a, b, c$, Table \ref{table_Admissible1} simplifies to Table \ref{table_Admissible2}, or in a vector form
\begin{equation}
u_{abc}^T=\left(u_a,u_b,u_c\right)\label{eq_uabc}
\end{equation}
where $u_k=\{-1,1\}$. Using the complex transformation \eqref{eq_ComplexTransf} with all the vectors in Table \ref{table_Admissible2} results in eight points on the complex plane, two of them at the origin (the \emph{zero vectors}) and six other combinations on the circumference with radius $2c$. Figure \ref{fig_2level} represents the complex values for the switch combinations on the circumference.

\begin{table}[htbp]
\centering
\begin{tabular}{||c||c|c|c|c|c|c|c|c||}
	\hline 
~ & $V_1$ & $V_2$  & $V_3$  & $V_4$ & $V_5$ & $V_6$ & $V_0^+$ & $V_0^-$\\ 
	\hline 
$u_a$ & $1$ & $1$ & $-1$ & $-1$ & $-1$ & $1$ & $1$ & $-1$\\ 
	\hline 
$u_b$ & $-1$ & $1$ & $1$ & $1$ & $-1$ & $-1$ & $1$ & $-1$\\ 
	\hline 
$u_c$ & $-1$ & $-1$ & $-1$ & $1$ & $1$ & $1$ & $1$ & $-1$\\
	\hline 
\end{tabular} \vspace{0.1cm}
\caption{Admissible vector combinations.}
\label{table_Admissible2}
\end{table}


The implementation of complex-valued sliding mode controllers in three-phase power converters consists of approximating the values resulting from 
\begin{equation}
\bm{u}=-2c\frac{\bm{\sigma}}{|\bm{\sigma}|},\label{eq_uPC}
\end{equation}
to a switch combination. The next Section presents different approaches to approximate \eqref{eq_uPC} in three-phase bridges.

\section{Sampled-time implementation methods} \label{sec_SampledMethods}

Many control algorithms for power electronic converters are implemented in a microprocessor running and providing the control signal with a fixed sampling time, $T_s$. The first approach, already used in \cite{DHB2022}, is to implement \eqref{eq_uPC} updating the control signal at each sample according to the control. Then, two alternatives are presented in the following sections.

\subsection{Sector-based Implementation} \label{sec_SectorImplementation}

\begin{figure}
\centering
\scalebox{0.8}{
\def\Pos{1.2}
\begin{circuitikz}
\draw[-latex] (-3,0) -- (3,0) node[anchor=north]  {Re($\bm{u}$)};
\draw[-latex] (0,-3) -- (0,3) node[anchor=east]  {Im($\bm{u}$)};
\fill[UPC!10] (0,0) -- (-30:2) arc (-30:30:2) ;
\fill[UPC!20] (0,0) -- (30:2) arc (30:90:2) ;
\fill[UPC!30] (0,0) -- (90:2) arc (90:150:2) ;
\fill[UPC!40] (0,0) -- (150:2) arc (150:210:2) ;
\fill[UPC!50] (0,0) -- (-150:2) arc (-150:-90:2) ;
\fill[UPC!60] (0,0) -- (-90:2) arc (-90:-30:2) ;
\draw[UPC!50!black] 
(0:\Pos) node {S1}
(60:\Pos) node {S2}
(120:\Pos) node {S3}
(180:\Pos) node {S4}
(-120:\Pos) node {S5}
(-60:\Pos) node {S6}
;
\draw[thick,UPC] (0,0) circle (2);
\filldraw[black] 
(0:2) circle (1.5pt) node[anchor=south west] {\footnotesize $V_1$}
(60:2) circle (1.5pt) node[anchor=south west] {\footnotesize $V_2$}
(120:2) circle (1.5pt) node[anchor=south east] {\footnotesize $V_3$}
(180:2) circle (1.5pt) node[anchor=south east] {\footnotesize $V_4$}
(-120:2) circle (1.5pt) node[anchor=north east] {\footnotesize $V_5$}
(-60:2) circle (1.5pt) node[anchor=north west] {\footnotesize $V_6$}
;
\filldraw[UPC!85!black] 
(120:2) circle (1.5pt) node[anchor=south east] {\footnotesize $V_3$}
;
\draw[thick,fill=red,red] (105:2) circle (1.5pt) node[anchor=south ] { $\bm{u}$};
\end{circuitikz}
}
\caption{Sectors defined by the Sector-based Implementation. The example (in red) shows a complex control value with $105^\circ$ that belongs to Sector S3, and the control switching vector, $V_3$,  $u_{abc}=(-1,1,-1)$, is applied (in blue).}\label{fig_SectorImplementation}
\end{figure}
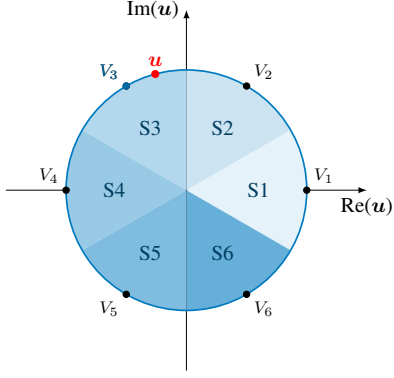

The Sector-based Implementation method (SbI) approximates the complex control value on the complex circle by the closest active vector. To this end, six zones are defined, as shown in Figure \ref{fig_SectorImplementation}. The switching policy is obtained according to Table \ref{table_SbI}. For example, if at a specific sample time, the angle of the complex control action $\bm{u}$ is $105^\circ$, this value belongs to Sector S3 and the switch combination corresponding to vector $V_3$,  $u_{abc}=(-1,1,-1)$, is applied.

\begin{table}[htbp]
\centering
\begin{tabular}{||c||c||c|c|c||}
	\hline 
$\angle \bm{u}$ & Sec. & $u_a$  & $u_b$  & $u_c$\\ 
	\hline 
$-30^\circ/30^\circ$ & S1 & $1$ & $-1$ & $-1$ \\ 
	\hline 
$30^\circ/90^\circ$ & S2 & $1$ & $1$ & $-1$ \\ 
	\hline 
$90^\circ/150^\circ$ & S3 & $-1$ & $1$ & $-1$ \\ 
	\hline 
$150^\circ/210^\circ$ & S4 & $-1$ & $1$ & $1$ \\ 
	\hline 
$210^\circ/270^\circ$ & S5 & $-1$ & $-1$ & $1$ \\ 
	\hline 
$270^\circ/330^\circ$ & S6 & $1$ & $-1$ & $1$ \\ 
	\hline 
\end{tabular} \vspace{0.1cm}
\caption{Sector and switch combination of the Sector-based Implementation.}
\label{table_SbI}
\end{table}

Notice that the Sector-based Implementation method does not use the zero vectors, defined in Table  \ref{table_Admissible2}.

%

\subsection{Complex Sliding Averaging} \label{sec_CSA}

\begin{figure}
\centering
\scalebox{0.8}{
\def\Pos{1.2}
\begin{circuitikz}
\draw[-latex] (-3,0) -- (3,0) node[anchor=north]  {Re($\bm{u}$)};
\draw[-latex] (0,-3) -- (0,3) node[anchor=east]  {Im($\bm{u}$)};
\fill[UPC!10] (0,0) -- (0:2) arc (0:60:2) ;
\fill[UPC!20] (0,0) -- (60:2) arc (60:120:2) ;
\fill[UPC!30] (0,0) -- (120:2) arc (120:180:2) ;
\fill[UPC!40] (0,0) -- (180:2) arc (-180:-120:2) ;
\fill[UPC!50] (0,0) -- (-120:2) arc (-120:-60:2) ;
\fill[UPC!60] (0,0) -- (-60:2) arc (-60:0:2) ;
\draw[UPC!50!black] 
(30:\Pos) node {S1}
(90:\Pos) node {S2}
(150:\Pos) node {S3}
(-150:\Pos) node {S4}
(-90:\Pos) node {S5}
(-30:\Pos) node {S6}
;
\draw[thick,UPC] (0,0) circle (2);
\filldraw[black] 
(0:2) circle (1.5pt) node[anchor=south west] {\footnotesize $V_1$}
(60:2) circle (1.5pt) node[anchor=south west] {\footnotesize $V_2$}
(120:2) circle (1.5pt) node[anchor=south east] {\footnotesize $V_3$}
(180:2) circle (1.5pt) node[anchor=south east] {\footnotesize $V_4$}
(-120:2) circle (1.5pt) node[anchor=north east] {\footnotesize $V_5$}
(-60:2) circle (1.5pt) node[anchor=north west] {\footnotesize $V_6$}
;
\draw[thick,fill=red,red] (105:2) circle (1.5pt) node[anchor=south ] { $\bm{u}$};
\draw[-latex,UPC!85!black] (0,0) -- (120:1.5);
\draw[-latex,UPC!85!black] (120:1.5) -- ++ (60:0.5) coordinate(uhat);
\filldraw[UPC!85!black] (uhat) circle(1.5pt) node[anchor=west ] { $\hat{\bm{u}}$};
\end{circuitikz}
}
\caption{Sectors defined by the Complex Sliding Averaging (CSA). The example (in red) shows a complex control value with $105^\circ$ that belongs to Sector S2, and the averaged equivalent complex vector is applied $\hat{\bm{u}}$ (in blue).
}\label{fig_CSA}
\end{figure}
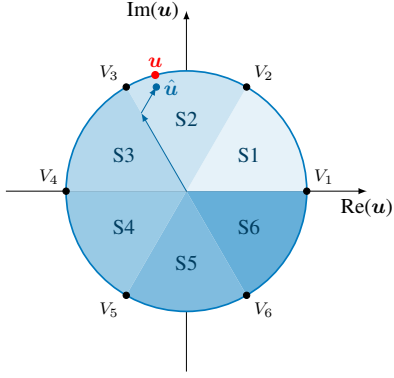

The Complex Sliding Averaging (CSA) consists of averaging the three-phase control action $u_{abc}$ between two control vectors. The complex plane is divided into six sectors, as shown in Figure \ref{fig_CSA}. Note that this sector definition differs from the one used by the SbI method proposed in Section \ref{sec_SectorImplementation}, and they are shifted $30^\circ$. The implementation technique defines a duty cycle, $d$, calculated as
\begin{equation}
d=\frac{\angle \bm{u}-\angle \bm{u}^-}{60^\circ},\label{eq_Duty}
\end{equation}
where $\angle \bm{u}^-$ depends on the sector where $\bm{u}$ belongs, see Table \ref{table_CSA}. The obtained value is compared with a carrier signal, providing the active time for two switch combinations $u_{abc}^+$ and $u_{abc}^-$ according to Table \ref{table_CSA}. 

\begin{table}[htbp]
\centering
\begin{tabular}{||c||c||c|c|c|c||}
	\hline 
$\angle \bm{u}$ & Sec. & $u_{abc}^+$  & $u_{abc}^-$ & $\angle \bm{u}^+$ & $\angle \bm{u}^-$\\ 
	\hline 
$0^\circ/60^\circ$ & S1 & $(1,1,-1)$ & $(1,-1,-1)$ & $60^\circ$ & $0^\circ$\\ 
	\hline 
$60^\circ/120^\circ$ & S2 & $(-1,1,-1)$ & $(1,1,-1)$ & $120^\circ$ & $60^\circ$\\ 
	\hline 
$120^\circ/180^\circ$ & S3 & $(-1,1,1)$ & $(-1,1,-1)$ & $180^\circ$ & $120^\circ$\\ 
	\hline 
$180^\circ/240^\circ$ & S4 & $(-1,-1,1)$ & $(-1,1,1)$ & $240^\circ$ & $180^\circ$\\ 
	\hline 
$240^\circ/300^\circ$ & S5 & $(1,-1,1)$ & $(-1,-1,1)$ & $300^\circ$ & $240^\circ$\\ 
	\hline 
$300^\circ/360^\circ$ & S6 & $(1,-1,-1)$ & $(1,-1,1)$ & $360^\circ$ & $300^\circ$\\ 
	\hline 
\end{tabular} \vspace{0.1cm}
\caption{Sectors and switch combinations of the Complex Sliding Averaging (CSA).}
\label{table_CSA}
\end{table}

\begin{figure}
\centering
\scalebox{0.8}{
\def\h{3.5} \def\Ts{3}
\def\da{0.7} \def\db{0.45} \def\dc{0.65}
\begin{circuitikz}
\draw[-latex] (0,0) -- ($\Ts*(3,0)+(0.5,0)$) node[anchor=north]  {$t$};
\draw[-latex] (0,0) -- ($\h*(0,1)+(0,0.5)$);

\draw[thick,UPC] (0,0)--++ ($\Ts*(0.5,0)+\h*(0,1)$) --++  ($\Ts*(0.5,0)+\h*(0,-1)$)
--++  ($\Ts*(0.5,0)+\h*(0,1)$) --++  ($\Ts*(0.5,0)+\h*(0,-1)$) 
--++  ($\Ts*(0.5,0)+\h*(0,1)$) --++  ($\Ts*(0.5,0)+\h*(0,-1)$) ;

\draw[thick,red] 
($\h*(0,\da)$) node[anchor=south west]  {$d_{p-1}$} -- ($\Ts*(1,0)+\h*(0,\da)$) 
($\Ts*(1,0)+\h*(0,\db)$) node[anchor=south west]  {$d_p$} -- ($\Ts*(2,0)+\h*(0,\db)$)
($\Ts*(2,0)+\h*(0,\dc)$) node[anchor=south west]  {$d_{p+1}$} -- ($\Ts*(3,0)+\h*(0,\dc)$)
 ;
\draw[dashed,UPC!85!black] 
($\da*\h*(0,1)+\da*\Ts/2*(1,0)$) -- ($(0,-1)+\da*\Ts/2*(1,0)$)
($\da*\h*(0,1)+\da*\Ts/2*(-1,0)+\Ts*(1,0)$) -- ($(0,-1)+\da*\Ts/2*(-1,0)+\Ts*(1,0)$)
($\db*\h*(0,1)+\db*\Ts/2*(1,0)+\Ts*(1,0)$) -- ($(0,-1)+\db*\Ts/2*(1,0)+\Ts*(1,0)$)
($\db*\h*(0,1)+\db*\Ts/2*(-1,0)+\Ts*(2,0)$) -- ($(0,-1)+\db*\Ts/2*(-1,0)+\Ts*(2,0)$)
($\dc*\h*(0,1)+\dc*\Ts/2*(1,0)+\Ts*(2,0)$) -- ($(0,-1)+\dc*\Ts/2*(1,0)+\Ts*(2,0)$)
($\dc*\h*(0,1)+\dc*\Ts/2*(-1,0)+\Ts*(3,0)$) -- ($(0,-1)+\dc*\Ts/2*(-1,0)+\Ts*(3,0)$)
;
\fill[UPC!60]
(0,0) rectangle ($(0,-1)+\da*\Ts/2*(1,0)+\Ts*(0,0)$) node[pos=0.5]  {\color{black}$\underbrace{u_{abc}^+}_{p-1}$}
($(0,0)+\da*\Ts/2*(-1,0)+\Ts*(1,0)$) rectangle ($(0,-1)+\Ts*(1,0)$) node[pos=0.5]  {\color{black}$\underbrace{u_{abc}^+}_{p-1}$}
($(0,0)+\Ts*(1,0)$) rectangle ($(0,-1)+\db*\Ts/2*(1,0)+\Ts*(1,0)$) node[pos=0.5]  {\color{black}$\underbrace{u_{abc}^+}_{p}$}
($(0,0)+\db*\Ts/2*(-1,0)+\Ts*(2,0)$) rectangle ($(0,-1)+\Ts*(2,0)$) node[pos=0.5]  {\color{black}$\underbrace{u_{abc}^+}_p$}
($(0,0)+\Ts*(2,0)$) rectangle ($(0,-1)+\dc*\Ts/2*(1,0)+\Ts*(2,0)$) node[pos=0.5]  {\color{black}$\underbrace{u_{abc}^+}_{p+1}$}
($(0,0)+\dc*\Ts/2*(-1,0)+\Ts*(3,0)$) rectangle ($(0,-1)+3*\Ts*(1,0)$) node[pos=0.5]  {\color{black}$\underbrace{u_{abc}^+}_{p+1}$}
;
\fill[UPC!40]
($(0,-1)+\da*\Ts/2*(1,0)+\Ts*(0,0)$) rectangle ($(0,0)+\da*\Ts/2*(-1,0)+\Ts*(1,0)$) node[pos=0.5]  {\color{black}$\underbrace{u_{abc}^-}_{p-1}$}
($(0,-1)+\db*\Ts/2*(1,0)+\Ts*(1,0)$) rectangle ($(0,0)+\db*\Ts/2*(-1,0)+\Ts*(2,0)$) node[pos=0.5]  {\color{black}$\underbrace{u_{abc}^-}_p$}
($(0,-1)+\dc*\Ts/2*(1,0)+\Ts*(2,0)$) rectangle ($(0,0)+\dc*\Ts/2*(-1,0)+\Ts*(3,0)$) node[pos=0.5]  {\color{black}$\underbrace{u_{abc}^-}_{p+1}$}
;
\draw[dotted,black]
($\Ts*(1,0)$) -- ++ ($\h*(0,1)+(0,0.5)$)
($\Ts*(1,0)$) -- ++ ($\h*(0,-0.29)-(0,0.0)$)
($\Ts*(2,0)$) -- ++ ($\h*(0,1)+(0,0.5)$)
($\Ts*(2,0)$) -- ++ ($\h*(0,-0.29)-(0,0.0)$)
;
\draw[black]
($\Ts*(1,0)$) -- ++ (0,-1)
($\Ts*(2,0)$) -- ++ (0,-1)
;
\end{circuitikz}
}
\caption{Complex Sliding Averaging (CSA), where $p$ refers to the iteration number.}\label{fig_CSA_carrier}
\end{figure}
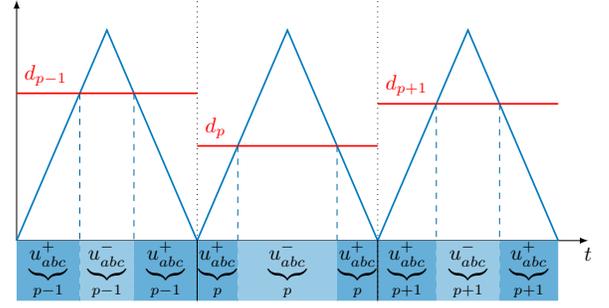

Figure \ref{fig_CSA_carrier} shows an example of three periods with duty cycles calculated at each sample time (in red) using a triangular pulse carrier function. The comparison of both signals provides the three-phase control vector to be applied. The instantaneous control signal in one period is
\begin{equation}\label{eq_uabct}
u_{abc}(t)=\left\{\begin{array}{ll}
u_{abc}^+&\mathrm{if~}t<t_0+\frac{d}{2}T_s\\
u_{abc}^-&\mathrm{if~}t_0+\frac{d}{2}T_s\leq t<t_0+\frac{2-d}{2}T_s\\
u_{abc}^+&\mathrm{if~}t\geq t_0+\frac{2-d}{2}T_s
\end{array}\right.
\end{equation}
where $t_0$ is the initial sampling time. The average of \eqref{eq_uabct} in one period is
\begin{equation}
\hat{u}_{abc}=\frac{1}{T_s}\int_{t_0}^{t_0+Ts}u_{abc}(\tau)\mathrm{d}\tau,
\end{equation}
resulting in
\begin{equation}
\hat{u}_{abc}=du_{abc}^++(1-d)u_{abc}^-.\label{eq_uabcCSA}
\end{equation}

However, the control action obtained from \eqref{eq_uabcCSA} with the duty definition used in \eqref{eq_Duty} does not provide the exact value of $\bm{u}$. Along the following lines, the deviation of the desired control action and the actual one is calculated.
The averaged equivalent complex control action applied with \eqref{eq_uabcCSA} can be obtained using the complex transformation \eqref{eq_ComplexTransf}, and results in
\begin{equation}
\hat{\bm{u}}=d\bm{u}^++(1-d)\bm{u}^-\label{eq_uhat}
\end{equation}
where $\bm{u}^+=\bm{T}u_{abc}^+$, $\bm{u}^-=\bm{T}u_{abc}^-$ are complex values. The average value in \eqref{eq_uhat} can be compared with the desired complex control action $\bm{u}$. The modulus of the deviation is
\begin{equation}
e_\text{mod}=\frac{|\bm{u}|-|\hat{\bm{u}}|}{|\bm{u}|},
\end{equation}
and since $|\bm{u}|=|\bm{u}^+|=|\bm{u}^-|$
\begin{align}
e_\text{mod}=&1-\left|d\cos(\angle \bm{u}^+)+(1-d)\cos(\angle \bm{u}^-)\right.\nonumber
\\&+\left.j(d\sin(\angle \bm{u}^+)+(1-d)\sin(\angle \bm{u}^-)\right|,\label{eq_emod}
\end{align}
where values $\angle \bm{u}^+$, $\angle \bm{u}^-$ are described in Table \ref{table_CSA}. The phase deviation is 
\begin{equation}
e_\text{ph}=\angle \bm{u}-\angle\hat{\bm{u}}.
\end{equation}
From \eqref{eq_uhat} and using \eqref{eq_Duty}, the phase deviation is described as a function of $d$ as
\begin{align}
e_\text{ph}=&60^\circ d +\angle \bm{u}^-\nonumber\\&-\arctan\left(\frac{d\sin(\angle \bm{u}^+)+(1-d)\sin(\angle \bm{u}^-)}{d\cos(\angle \bm{u}^+)+(1-d)\cos(\angle \bm{u}^-)}\right).\label{eq_eph}
\end{align}

\begin{figure}
\centering
\includegraphics[width=8.5cm]{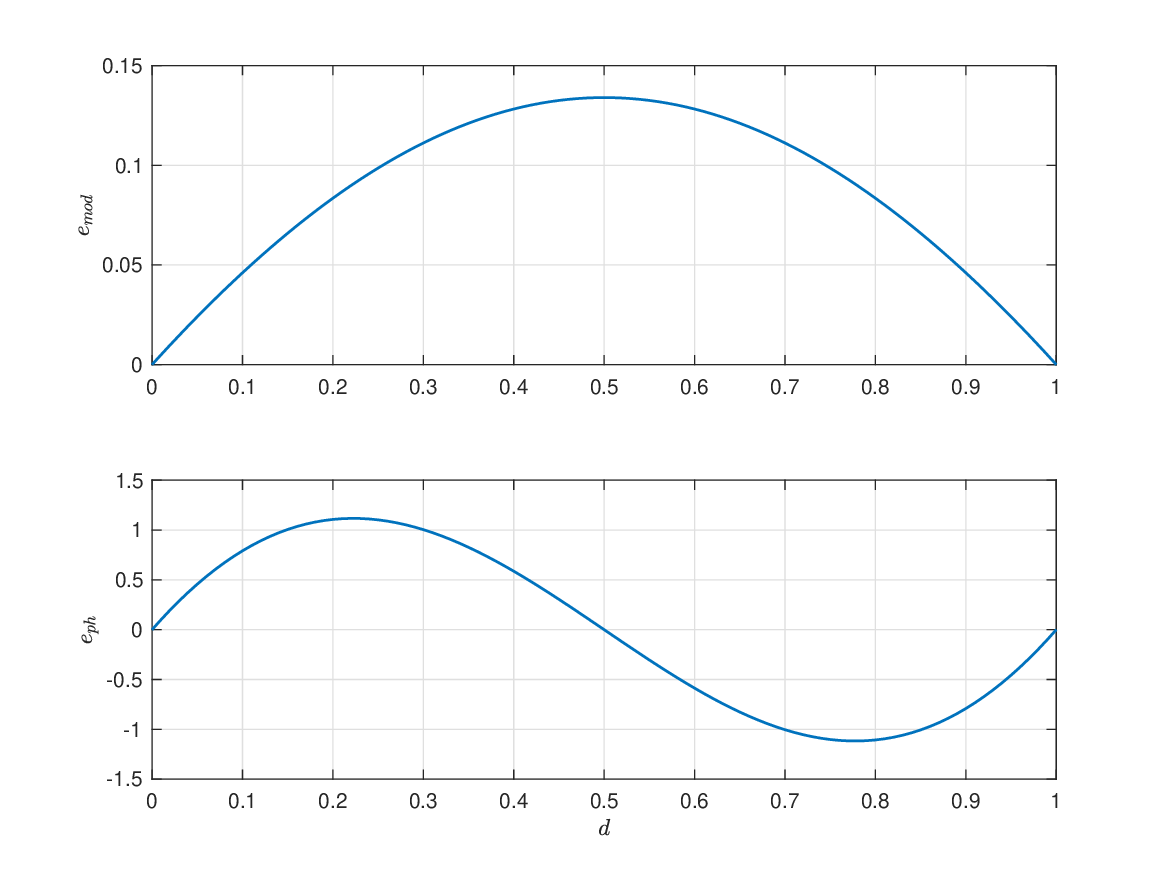}
\caption{Modulus (top) and phase (bottom) deviation functions when applying the CSA.}
\label{fig_ErrorDuty}
\end{figure}

The deviation functions \eqref{eq_emod} and \eqref{eq_eph} are the same for all sectors and are shown in Figure \ref{fig_ErrorDuty}. The maximum deviation can be calculated by comparing the average with the desired value. Let us analyse the case for Sector S1, then $\bm{u}^+=2c\left(\frac{1}{2}+j\frac{\sqrt{3}}{2}\right)$, $\bm{u}^-=2c$, and the averaged complex control action can be written as
\begin{equation}
\hat{\bm{u}}=2c\left(1-d\frac{1}{2}+jd\frac{\sqrt{3}}{2}\right).
\end{equation}
Then, the deviation modulus is simplified to
\begin{equation}
e_\text{mod}=1-\sqrt{d^2-d+1}.
\end{equation}
with a maximum value
\begin{equation}
e_\text{mod}^\text{max}=1-\frac{\sqrt{3}}{2}\approx0.134\label{eq_maxdeviation}
\end{equation}
for $d=0.5$. The phase deviation results in\footnote{The output of $\arctan$ function should be taken in degrees.}
\begin{equation}
e_\text{ph}=60^\circ d+\arctan\left(\sqrt{3}\frac{d}{d-2}\right)
\end{equation}
with a maximum absolute deviation $e_\text{ph}^\text{max}=1.117^\circ$ for $d=0.222$ and $d=0.777$.

Figure \ref{fig_CSA} shows the same example of a control value with $105^\circ$ used in the previous Section. The red dot indicates the desired control value is now placed in Sector S2. From \eqref{eq_Duty}, the duty becomes $d=0.75$, and the control action \eqref{eq_uabcCSA} mapped to the complex plane results in the averaged complex control $\hat{\bm{u}}$ (in blue). As expected, the $\bm{u}$ and $\hat{\bm{u}}$ values are different.


\subsection{Zero-vector Complex Sliding Averaging (zCSA)} \label{sec_zCSA}

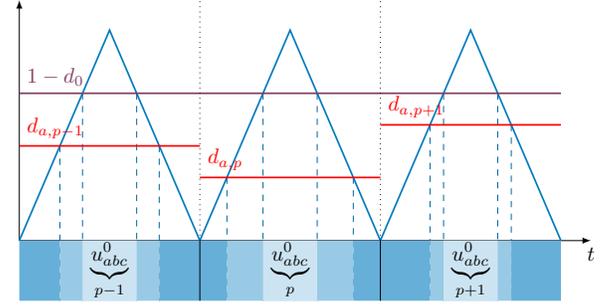
\begin{figure}
\centering
\scalebox{0.8}{
\def\h{3.5} \def\Ts{3.0}
\def\do{0.7}
\def\da{0.45} \def\db{0.3} \def\dc{0.55}
\begin{circuitikz}
\draw[-latex] (0,0) -- ($\Ts*(3,0)+(0.5,0)$) node[anchor=north]  {$t$};
\draw[-latex] (0,0) -- ($\h*(0,1)+(0,0.5)$);

\draw[thick,UPC] (0,0)--++ ($\Ts*(0.5,0)+\h*(0,1)$) --++  ($\Ts*(0.5,0)+\h*(0,-1)$)
--++  ($\Ts*(0.5,0)+\h*(0,1)$) --++  ($\Ts*(0.5,0)+\h*(0,-1)$) 
--++  ($\Ts*(0.5,0)+\h*(0,1)$) --++  ($\Ts*(0.5,0)+\h*(0,-1)$) ;

\draw[dashed,UPC!85!black] 
($\da*\h*(0,1)+\da*\Ts/2*(1,0)$) -- ($(0,-1)+\da*\Ts/2*(1,0)$)
($\da*\h*(0,1)+\da*\Ts/2*(-1,0)+\Ts*(1,0)$) -- ($(0,-1)+\da*\Ts/2*(-1,0)+\Ts*(1,0)$)
($\db*\h*(0,1)+\db*\Ts/2*(1,0)+\Ts*(1,0)$) -- ($(0,-1)+\db*\Ts/2*(1,0)+\Ts*(1,0)$)
($\db*\h*(0,1)+\db*\Ts/2*(-1,0)+\Ts*(2,0)$) -- ($(0,-1)+\db*\Ts/2*(-1,0)+\Ts*(2,0)$)
($\dc*\h*(0,1)+\dc*\Ts/2*(1,0)+\Ts*(2,0)$) -- ($(0,-1)+\dc*\Ts/2*(1,0)+\Ts*(2,0)$)
($\dc*\h*(0,1)+\dc*\Ts/2*(-1,0)+\Ts*(3,0)$) -- ($(0,-1)+\dc*\Ts/2*(-1,0)+\Ts*(3,0)$)
;
\draw[dashed,UPC!85!black] 
($\do*\h*(0,1)+\do*\Ts/2*(1,0)$) -- ($(0,-1)+\do*\Ts/2*(1,0)$)
($\do*\h*(0,1)+\do*\Ts/2*(-1,0)+\Ts*(1,0)$) -- ($(0,-1)+\do*\Ts/2*(-1,0)+\Ts*(1,0)$)
($\do*\h*(0,1)+\do*\Ts/2*(1,0)+\Ts*(1,0)$) -- ($(0,-1)+\do*\Ts/2*(1,0)+\Ts*(1,0)$)
($\do*\h*(0,1)+\do*\Ts/2*(-1,0)+\Ts*(2,0)$) -- ($(0,-1)+\do*\Ts/2*(-1,0)+\Ts*(2,0)$)
($\do*\h*(0,1)+\do*\Ts/2*(1,0)+\Ts*(2,0)$) -- ($(0,-1)+\do*\Ts/2*(1,0)+\Ts*(2,0)$)
($\do*\h*(0,1)+\do*\Ts/2*(-1,0)+\Ts*(3,0)$) -- ($(0,-1)+\do*\Ts/2*(-1,0)+\Ts*(3,0)$)
;

\draw[thick,red!50!UPC] 
($\h*(0,\do)$) node[anchor=south west]  {$1-d_0$} -- ($\Ts*(3,0)+\h*(0,\do)$)
;
\draw[thick,red] 
($\h*(0,\da)$) node[anchor=south west]  {$d_{a,p-1}$} -- ($\Ts*(1,0)+\h*(0,\da)$) 
($\Ts*(1,0)+\h*(0,\db)$) node[anchor=south west]  {$d_{a,p}$} -- ($\Ts*(2,0)+\h*(0,\db)$)
($\Ts*(2,0)+\h*(0,\dc)$) node[anchor=south west]  {$d_{a,p+1}$} -- ($\Ts*(3,0)+\h*(0,\dc)$)
 ;
\fill[UPC!60]
(0,0) rectangle ($(0,-1)+\da*\Ts/2*(1,0)+\Ts*(0,0)$) 
($(0,0)+\da*\Ts/2*(-1,0)+\Ts*(1,0)$) rectangle ($(0,-1)+\db*\Ts/2*(1,0)+\Ts*(1,0)$) 
($(0,0)+\db*\Ts/2*(-1,0)+\Ts*(2,0)$) rectangle ($(0,-1)+\dc*\Ts/2*(1,0)+\Ts*(2,0)$) 
($(0,0)+\dc*\Ts/2*(-1,0)+\Ts*(3,0)$) rectangle ($(0,-1)+3*\Ts*(1,0)$) 
;
\fill[UPC!40]
($(0,-1)+\da*\Ts/2*(1,0)+\Ts*(0,0)$) rectangle ($(0,0)+\da*\Ts/2*(-1,0)+\Ts*(1,0)$) 
($(0,-1)+\db*\Ts/2*(1,0)+\Ts*(1,0)$) rectangle ($(0,0)+\db*\Ts/2*(-1,0)+\Ts*(2,0)$) 
($(0,-1)+\dc*\Ts/2*(1,0)+\Ts*(2,0)$) rectangle ($(0,0)+\dc*\Ts/2*(-1,0)+\Ts*(3,0)$) 
;
\fill[UPC!20]
($(0,-1)+\do*\Ts/2*(1,0)+\Ts*(0,0)$) rectangle ($(0,0)+\do*\Ts/2*(-1,0)+\Ts*(1,0)$) node[pos=0.5]  {\color{black}$\underbrace{u_{abc}^0}_{p-1}$}
($(0,-1)+\do*\Ts/2*(1,0)+\Ts*(1,0)$) rectangle ($(0,0)+\do*\Ts/2*(-1,0)+\Ts*(2,0)$) node[pos=0.5]  {\color{black}$\underbrace{u_{abc}^0}_p$}
($(0,-1)+\do*\Ts/2*(1,0)+\Ts*(2,0)$) rectangle ($(0,0)+\do*\Ts/2*(-1,0)+\Ts*(3,0)$) node[pos=0.5]  {\color{black}$\underbrace{u_{abc}^0}_{p+1}$}
;
\draw[dotted,black]
($\Ts*(1,0)$) -- ++ ($\h*(0,1)+(0,0.5)$)
($\Ts*(2,0)$) -- ++ ($\h*(0,1)+(0,0.5)$)
;
\draw[black]
($\Ts*(1,0)$) -- ++ (0,-1)
($\Ts*(2,0)$) -- ++ (0,-1)
;
\end{circuitikz}
}

\caption{Zero-vector Complex Sliding Averaging (zCSA).}\label{fig_zCSA_carrier}
\end{figure}

The CSA method presented in Section \ref{sec_CSA} does not apply zero vectors. However, zero vectors can be included, defining the control signal as 
\begin{equation}\label{eq_uabct2}
u_{abc}(t)=\left\{\begin{array}{ll}
u_{abc}^+&\mathrm{if~}t<t_0+\frac{d_a}{2}T_s\\
u_{abc}^-&\mathrm{if~}t_0+\frac{d_a}{2}T_s\leq t<t_0+\frac{1-d_0}{2}T_s\\
u_{abc}^0&\mathrm{if~}t_0+\frac{1-d_0}{2}T_s\leq t<t_0+\frac{1+d_0}{2}T_s\\
u_{abc}^-&\mathrm{if~}t_0+\frac{1+d_0}{2}T_s\leq t<t_0+\frac{2-d_a}{2}T_s\\
u_{abc}^+&\mathrm{if~}t\geq t_0+\frac{2-d_a}{2}T_s
\end{array}\right.
\end{equation}
where $d_0$ is the \emph{zero duty cycle} (the interval time in a sample period where zero vectors are applied), defined as $0\leq d_0<1$, and the applied duty cycle is modified as follows
\begin{equation}
d_a=(1-d_0)d=(1-d_0)\frac{\angle u-\angle u^-}{60^\circ}.\label{eq_DutyA}
\end{equation}
Figure \ref{fig_zCSA_carrier} shows the switching policy when zero vectors are applied, where $u_{abc}^0$ is chosen between $(1,1,1)$ or $(-1,-1,-1)$ such that only one phase switch is changed with respect to the active vector $u_{abc}^-$.


Averaging \eqref{eq_uabct2} over one period, one gets
\begin{equation}
\hat{u}_{abc}=d_0u^0_{abc}+d_au^+_{abc}+(1-d_0-d_a)u^-_{abc}.\label{eq_uabczCSA}
\end{equation}
Using the complex transformation \eqref{eq_ComplexTransf} and the applied duty cycle definition in \eqref{eq_DutyA}, the averaged equivalent complex control action applied results in\footnote{Note that, from $\bm{T}u_{abc}^0$, one has $\bm{u}^0=0$.}
\begin{equation}
\hat{\bm{u}}=(1-d_0)\left(d\bm{u}^++(1-d)\bm{u}^-\right),\label{eq_uhatA}
\end{equation}
which reveals that the phase of the averaged applied control remains, but the modulus is modified with a gain $1-d_0$. From Equation \eqref{eq_uhatA}, the use of zero vectors effectively reduces the modulus (gain) of the control action, i.e., is equivalent to 
\begin{equation}
\bm{u}=-2c(1-d_0)\frac{\bm{\sigma}}{|\bm{\sigma}|}.\label{eq_uPC_Zero}
\end{equation}
From the condition in  \eqref{eq_cond}, and using \eqref{eq_uPC_Zero}, 
\begin{equation}
2c(1-d_0)>|\bm{u}_{eq}|
\end{equation}
and the maximum zero duty cycle that ensures the existence of the sliding motion is
\begin{equation}
d_0^\mathrm{max}<1-\frac{1}{2c}|\bm{u}_{eq}|.\label{eq_d0max}
\end{equation}
Note that, for a constant $d_0$, the maximum deviation value obtained in \eqref{eq_maxdeviation} must still be considered.

The attractiveness of using zero vectors is that by reducing the effective gain of the sliding mode controller, the ripple of the output variables is reduced, and the performance of the controller improves. See the performance indexes obtained in the application example in Section \ref{sec_example}.


\section{Example: Complex-valued SMC of a voltage source inverter}\label{sec_example}

\subsection{Complex-valued description of a voltage source inverter}

\begin{figure}
\centering
\scalebox{0.75}{\begin{circuitikz}

\draw (0,-5) to [american voltage source,invert,l=$V_{dc}$]++(0,2.5) coordinate (auxdc)
	to [american voltage source,invert,l=$V_{dc}$]++(0,2.5)
	(auxdc) to [short,*-]++(-0.5,0) node [ground] {};
\draw (0,0) to [short,-*]++(1.5,0) coordinate (aux1)
	to [short,-*]++(1.5,0) coordinate (aux2)
	to [short]++(1.5,0) coordinate (aux3);

\draw (aux1) to [short]++(0,-0.25) node[nigbt,bodydiode, anchor=C](Sa){} (Sa.B) node[anchor=south] {$u_a$}
(Sa.E) to [short,-*]++(0,-0.25) coordinate(outA) to [short]++(0,-1.75)
to [short]++(0,-0.25) node[nigbt,bodydiode, anchor=C](SaN){} (SaN.B) node[anchor=south] {$\bar{u}_a$}
(SaN.E) to [short,-*]++(0,-0.25);

\draw (aux2) to [short]++(0,-0.25) node[nigbt,bodydiode, anchor=C](Sb){} (Sb.B) node[anchor=south] {$u_b$}
(Sb.E) to [short,-*]++(0,-1) coordinate(outB) to [short]++(0,-1.0)
to [short]++(0,-0.25) node[nigbt,bodydiode, anchor=C](SbN){} (SbN.B) node[anchor=south] {$\bar{u}_b$}
(SbN.E) to [short,-*]++(0,-0.25);

\draw (aux3) to [short]++(0,-0.25) node[nigbt,bodydiode, anchor=C](Sc){} (Sc.B) node[anchor=south] {$u_c$}
(Sc.E) to [short,-*]++(0,-1.75) coordinate(outC) to [short]++(0,-0.25)
to [short]++(0,-0.25) node[nigbt,bodydiode, anchor=C](ScN){} (ScN.B) node[anchor=south] {$\bar{u}_c$}
(ScN.E) to [short]++(0,-0.25)
to [short]++(-4.5,0) 
;

\draw (outA) to [short,-o]++(3.5,0) node[anchor=south] {$v_a'$} coordinate (va)
	(outB) to [short,-o]++(2,0) node[anchor=south] {$v_b'$} coordinate (vb)
	(outC) to [short,-o]++(0.5,0) node[anchor=south] {$v_c'$} coordinate (vc);
	
\draw (vb) to [short,o-] ++(0.25,0)
	to [L,l=$L$]++(1.5,0) to [short]++(1,0) node[anchor=south] {$v_b$} coordinate (vCb)  to [short]++(1.5,0)  to [R,l=$R_{L}$]++(1.25,0) coordinate (RLo)
	(va) to [short,o-] ++(0.25,0) 
	to [L,l=$L$]++(1.5,0) node[anchor=south] {$v_a$} coordinate (vCa) to [short]++(2.5,0) coordinate  (Load) to [R,l=$R_{L}$]++(1.25,0) to [short,-*] (RLo)
	(vc) to [short,o-] ++(0.25,0) 
	to [L,l=$L$]++(1.5,0) to [short]++(2,0) node[anchor=south] {$v_c$} coordinate (vCc)  to [short]++(0.5,0) to [R,l=$R_{L}$]++(1.25,0) to [short,-*] (RLo)
;

\draw (vCb) to [short,*-]++(0,-1.25) to [C,l_=$C$,-*]++(0,-1) node[anchor=north] {$v_o$}  coordinate (vo)
	(vCa) to [short,*-]++(0,-2) to [C,l_=$C$]++(0,-1) to [short] (vo)
	(vCc) to [short,*-]++(0,-0.5) to [C,l_=$C$]++(0,-1) to [short] (vo)
;
\begin{pgfonlayer}{background}
\draw[fill=UPC!20,draw=none,rounded corners=1.5pt]  ($(Load)+(-0.0,0.75)$) rectangle++(1.5,-2.75) node[anchor=north  east]{\color{UPC}Load};
\end{pgfonlayer}

\end{circuitikz}}
\caption{Three-phase voltage source inverter.}\label{fig_3PhaseVSC}
\end{figure}
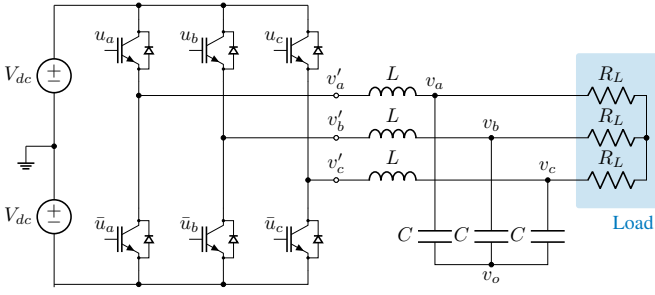

\begin{figure*}
\centering
\captionsetup[subfloat]{labelfont=scriptsize,textfont=scriptsize}    
\subfloat[Sector-based Implementation]{\includegraphics[width = 0.33\textwidth]{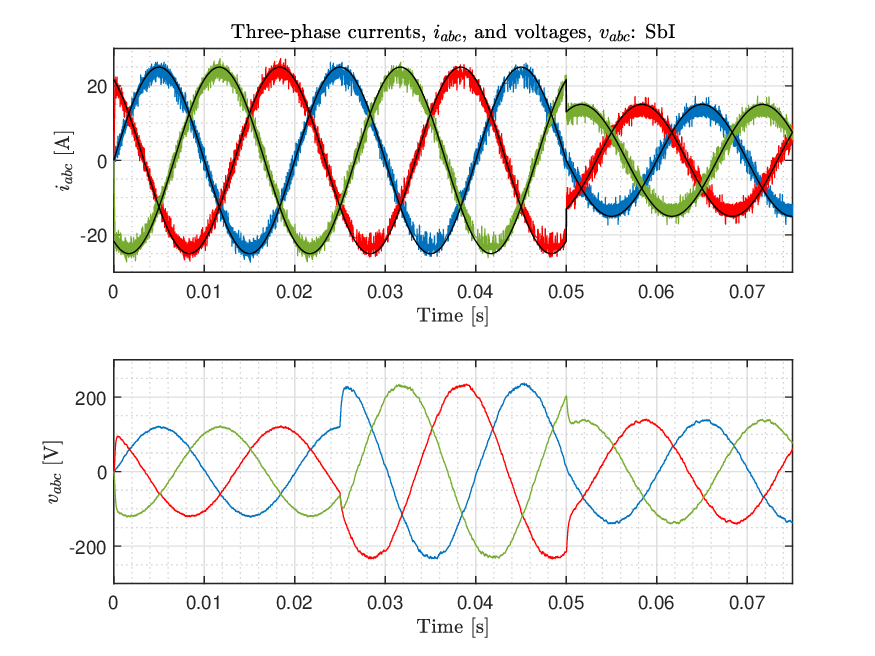}}
\subfloat[Complex Sliding Averaging]{\includegraphics[width = 0.33\textwidth]{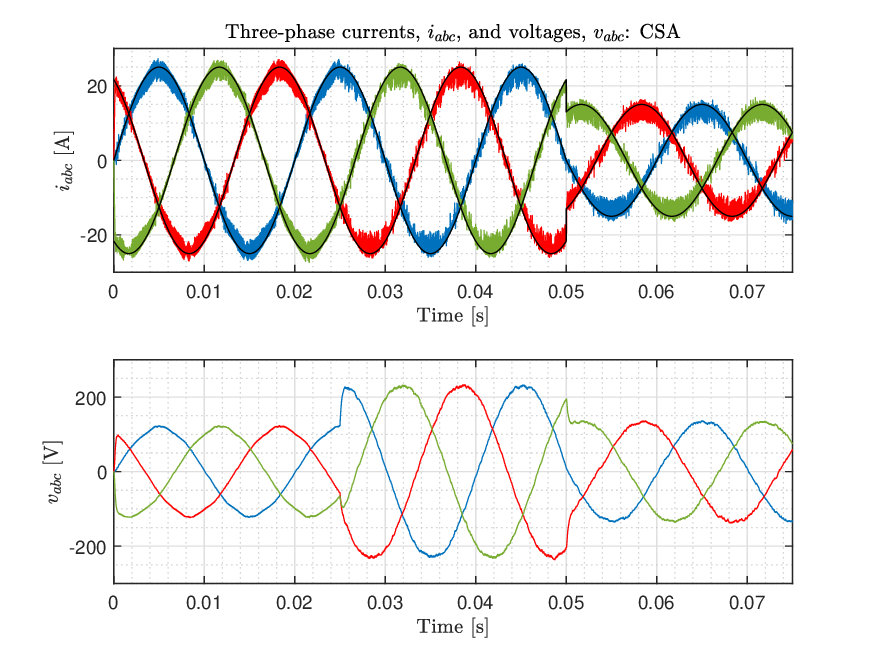}}
\subfloat[Zero-vector Complex Sliding Averaging]{\includegraphics[width = 0.33\textwidth]{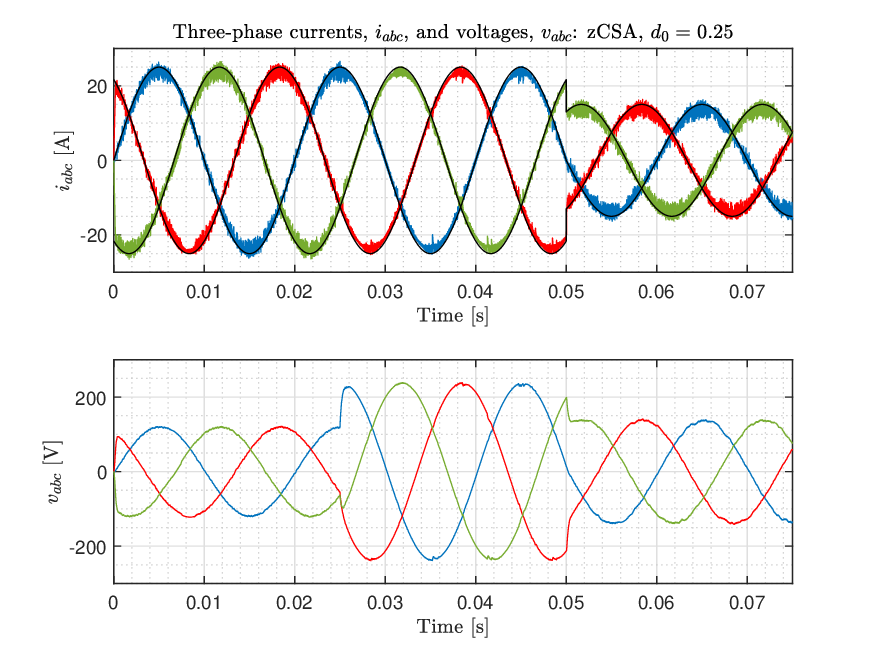}}
\caption{Simulation results: Three-phase currents (top) and three-phase voltages (bottom) for the different implementation methods.}
\label{fig_sims_transient}
\end{figure*}

This section uses a voltage source inverter (VSI) composed of the three-phase bridge connected to an LC filter and a resistive load to illustrate the control design proposed in Section \ref{sec_cSMCrevisited}, see the electrical scheme in Figure \ref{fig_3PhaseVSC}. The system is described by the space-state equations
\begin{align}
L\frac{\mathrm{d}i_{abc}}{\mathrm{d}t}=&-ri_{abc}-v_{abc}+V_{dc}u_{abc}-v_o,\\
C\frac{\mathrm{d}v_{abc}}{\mathrm{d}t}=&i_{abc}-\frac{1}{R_L}v_{abc},
\end{align}
where $\pm V_{dc}$ represents the high/low dc-voltage, $v_o$ is the voltage at the center of the capacitors, $i_{abc}^T=(i_a,i_b,i_c)\in\mathbb{R}^3$ are the inductor currents, $v_{abc}^T=(v_a,v_b,v_c)\in\mathbb{R}^3$ are the capacitor voltages, $u_{abc}^T=(u_a,u_b,u_c)\in\mathbb{R}^3$ are the switching control actions, with $u_k=\{-1,1\}$,  $L, C, r$ are the inductance, capacitance and resistance inductor losses, respectively, and $R_L$ are the load resistances. The control goal is tracking a balanced set of current references
\begin{equation}
i_{abc}^\mathrm{ref}(t)=I^\mathrm{ref}\begin{pmatrix}\cos(\omega t)\\\cos(\omega t-\frac{2\pi}{3})\\\cos(\omega t+\frac{2\pi}{3})\end{pmatrix},\label{eq_vref}
\end{equation}
where $I^\mathrm{ref}$ and $\omega$ are the constant current amplitude and frequency.
 
Using the complex transformation in \eqref{eq_ComplexTransf} with $c=\frac{2}{3}$ (amplitude preserving) and assuming balanced conditions (with null homopolar component), the VSI dynamics can be written as
\begin{align}
L\frac{\mathrm{d}\bm{i}}{\mathrm{d}t}&=-r\bm{i}-\bm{v}+V_{dc}\bm{u},\label{eq_di}\\
C\frac{\mathrm{d}\bm{v}}{\mathrm{d}t}&=\bm{i}-\frac{1}{R_L}\bm{v},\label{eq_dv}
\end{align}
where $\bm{i},\bm{v},\bm{u}\in\mathbb{C}$. The reference \eqref{eq_vref} becomes the complex-valued signal $\bm{i}^\mathrm{ref}=I^\mathrm{ref}e^{j\omega t}$.

\subsection{Complex-valued SMC design}

The design of the complex-valued sliding mode controller follows the steps described in Section \ref{sec_cSMCrevisited}. The switching surface is selected as 
\begin{equation}
\bm{\sigma}=\bm{i}-\bm{i}^\mathrm{ref}.\label{eq_sigma}
\end{equation}

\begin{enumerate}
\item \emph{Equivalent control.} Defining the state vector as $\bm{z}=(\bm{i},\bm{v})^T$, complex-valued dynamics \eqref{eq_di}-\eqref{eq_dv}, can be written under the form \eqref{eq_sys} with
\begin{equation}
f(\bm{z})=\begin{pmatrix}-\frac{1}{L}(r\bm{i}+\bm{v})\\\frac{1}{C}(\bm{i}-\frac{1}{R_L}\bm{v})\end{pmatrix},\quad
g(\bm{z})=\begin{pmatrix}\frac{V_{dc}}{L}\\0\end{pmatrix}.\label{eq_fg}
\end{equation}
From \eqref{eq_sigma} and \eqref{eq_fg} one gets
\begin{equation}
\frac{\partial \bm{\sigma}}{\partial \bm{z}}g(\bm{z})=\frac{V_{dc}}{L},\label{eq_dsg}
\end{equation}
and the transversality condition is fulfilled. Moreover,
\begin{equation}
\frac{\partial\bm{\sigma}}{\partial \bm{z}}f(\bm{z})=-\frac{1}{L}\left(r\bm{i}+\bm{v}\right)
\end{equation}
and $\frac{\partial \bm{\sigma}}{\partial t}=-j\omega\bm{i}^\mathrm{ref}$. From \eqref{eq_ueq}, the equivalent control results in
\begin{equation}
\bm{u}_{eq}=\frac{1}{V_{dc}}\left(r\bm{i}+\bm{v}+j\omega L\bm{i}^\mathrm{ref}\right).\label{eq_ueqExample}
\end{equation}

\item \emph{Ideal sliding dynamics.}  Assuming $\bm{\sigma}=0$, implies $\bm{i}=\bm{i}^\mathrm{ref}$. Replacing the current and $\bm{u}=\bm{u}_{eq}$ in the dynamics \eqref{eq_di}-\eqref{eq_dv} one gets $\frac{\mathrm{d}\bm{i}}{\mathrm{d}t}=0$ and
\begin{equation}
C\frac{\mathrm{d}\bm{v}}{\mathrm{d}t}=-\frac{1}{R_L}\bm{v}+\bm{i}^\mathrm{ref},
\end{equation}
which implies that the voltage stabilizes at $\bm{v}=R_L\bm{i}^\mathrm{ref}$.

\item \emph{Switching control law.} 
The sliding motion on $\bm{\sigma}=0$ is guaranteed if the condition \eqref{eq_cond} holds. From the control switching action for the three-phase bridge in \eqref{eq_uPC}, $\bm{k}=2c$ and $\delta_{k}=0$. Moreover, from \eqref{eq_dsg}, $\delta_{\sigma_g}=0$. Then, with \eqref{eq_ueqExample}, condition \eqref{eq_cond} results in
\begin{equation}
V_{dc}>\frac{3}{4}\left|r\bm{i}+\bm{v}+j\omega L\bm{i}^\mathrm{ref}\right|.
\end{equation}
The inequality above indicates the required DC voltage that guarantees the sliding motion.
\end{enumerate}

\subsection{Numerical simulations} \label{sec_SimsSampledTime}

Numerical simulations using Matlab/Simulink have been carried out to evaluate and compare the implementation strategies presented in the paper. The VSI parameters are $L=2~\mathrm{mH}$, $C=20~\mu\mathrm{F}$, $r=2~\mathrm{m}\Omega$, and $V_{dc}=300~\mathrm{V}$. The sampling frequency has been set to $50~\mathrm{kHz}$, and a delay of one switching period has been added to emulate the final implementation. Initially, the current reference has been set to $I^\mathrm{ref}=25~\mathrm{A}$ with $50~\mathrm{Hz}$ frequency, i.e., $\omega=100\pi~\mathrm{rad/s}$, with a load of $R_L=5~\Omega$. The test consisted of a load change to $R_L=10~\Omega$ at $t=25~\mathrm{ms}$, and a reference change to $I^\mathrm{ref}=15~\mathrm{A}$ at $t=50~\mathrm{ms}$. 

According to the parameters, from \eqref{eq_d0max} and \eqref{eq_ueqExample}, the maximum value of $d_0$ ensuring the sliding motion is $d_0^\mathrm{max}=0.3736$. For this reason, in the simulation and experimental tests, the zero duty cycle was set $d_0 = 0.25$ for a safety margin.

\begin{figure*}
\centering
\captionsetup[subfloat]{labelfont=scriptsize,textfont=scriptsize}    
\subfloat[Sector-based Implementation]{\includegraphics[width = 0.33\textwidth]{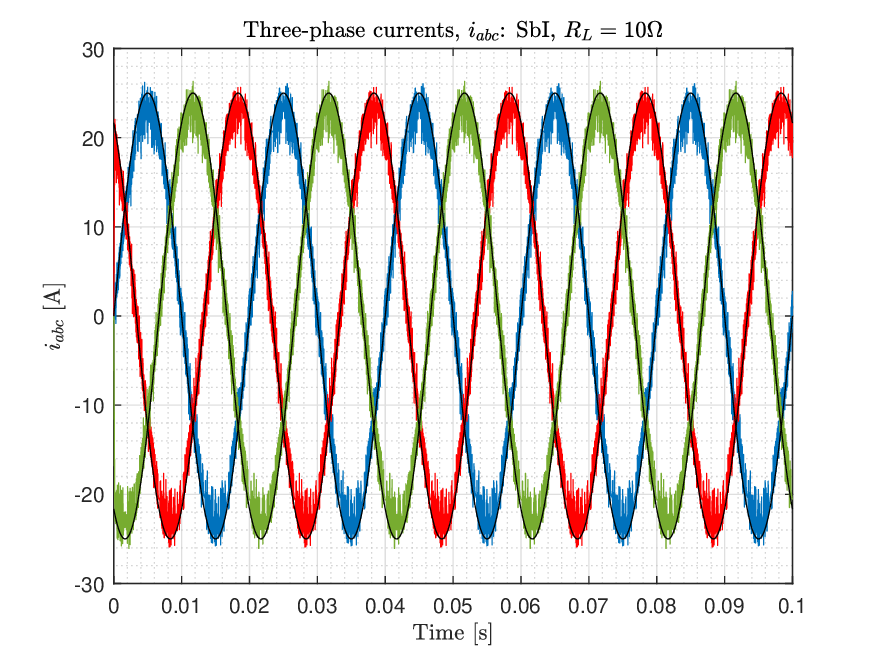}}
\subfloat[Complex Sliding Averaging]{\includegraphics[width = 0.33\textwidth]{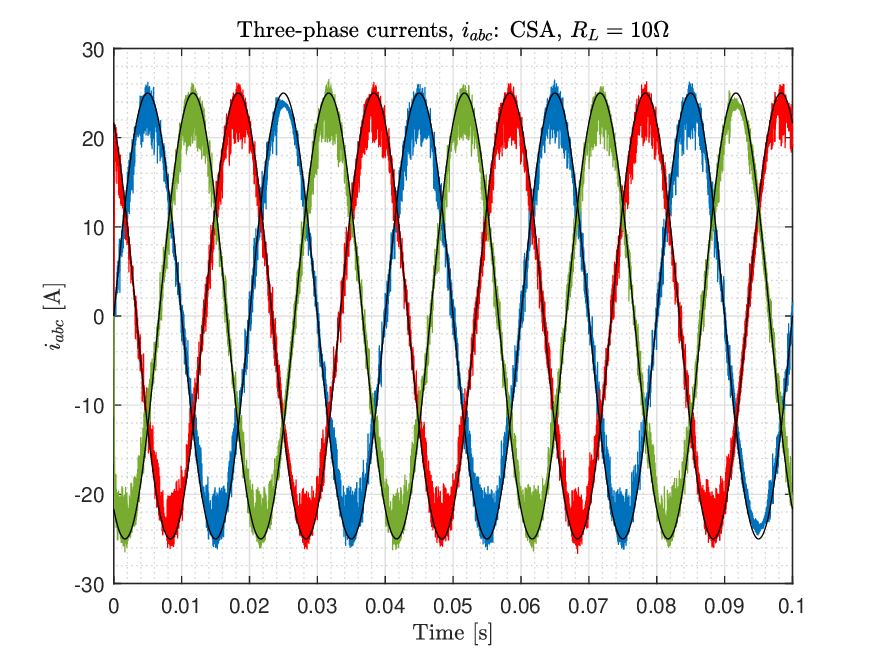}}
\subfloat[Zero-vector Complex Sliding Averaging]{\includegraphics[width = 0.33\textwidth]{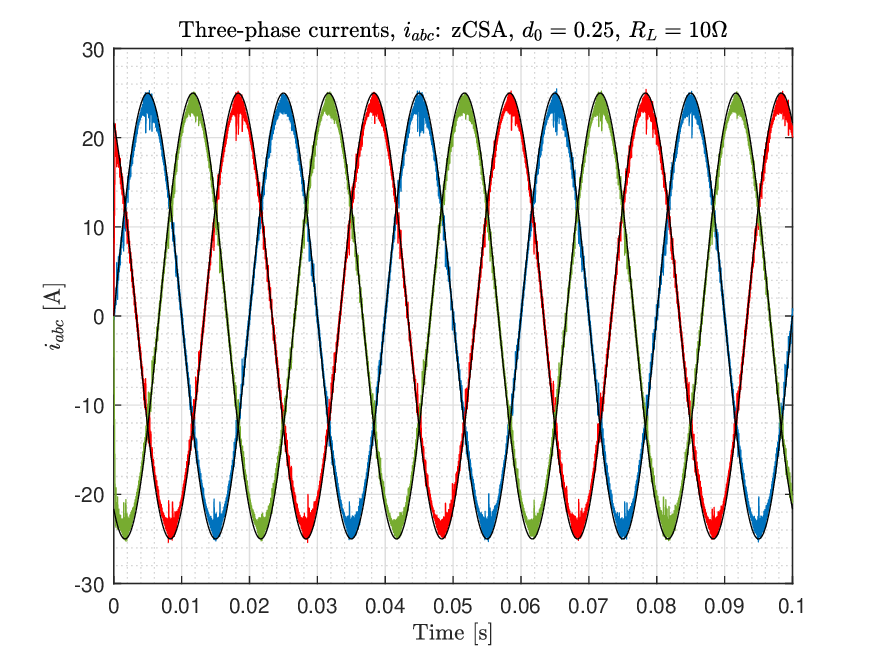}}
\caption{Simulation results: Three-phase currents for the different implementation methods.}
\label{fig_sims_i}
\end{figure*}

\begin{figure*}
\centering
\captionsetup[subfloat]{labelfont=scriptsize,textfont=scriptsize}    
\subfloat[Sector-based Implementation]{\includegraphics[width = 0.33\textwidth]{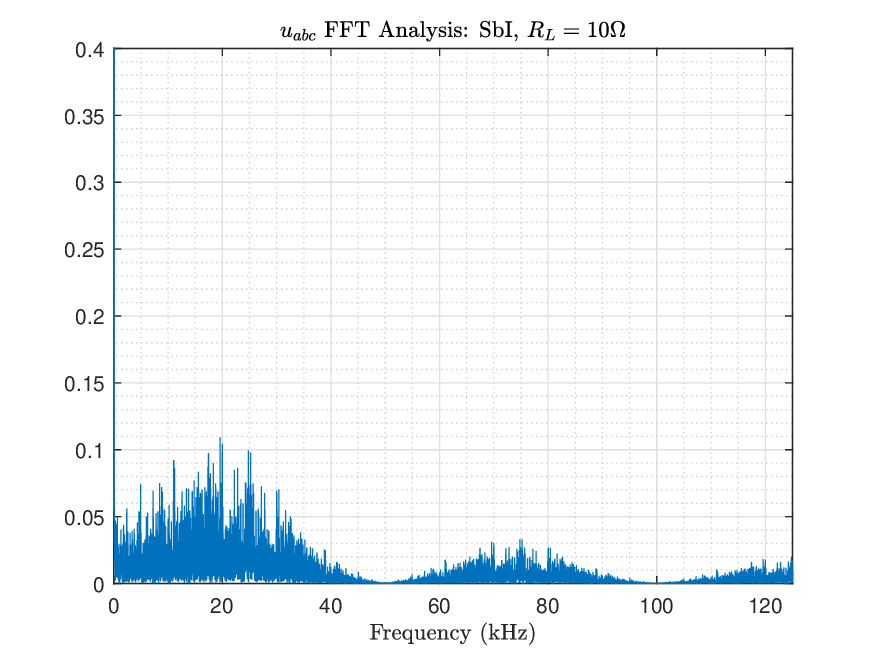}}
\subfloat[Complex Sliding Averaging]{\includegraphics[width = 0.33\textwidth]{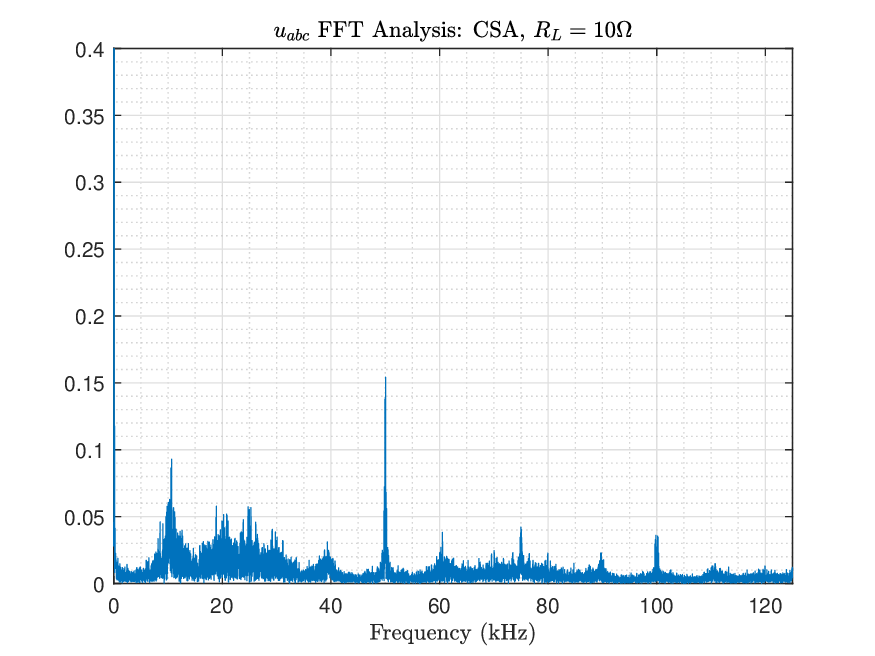}}
\subfloat[Zero-vector Complex Sliding Averaging]{\includegraphics[width = 0.33\textwidth]{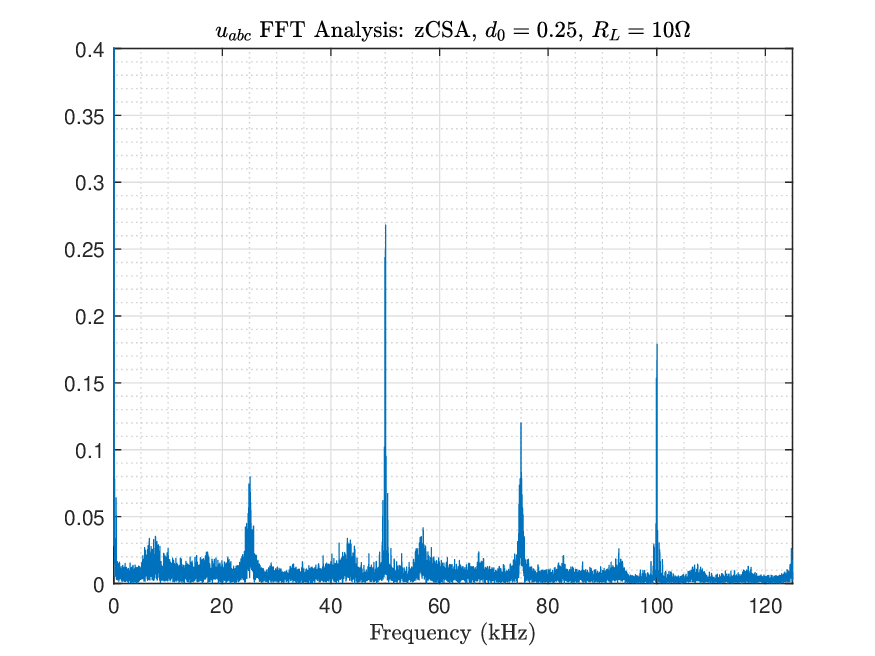}}
\caption{Simulation results: FFT of the switching control action, $u_{abc}$,  for the different implementation methods.}
\label{fig_sims_fft}
\end{figure*}

\begin{figure*}
\centering
\captionsetup[subfloat]{labelfont=scriptsize,textfont=scriptsize}    
\subfloat[Sector-based Implementation]{\includegraphics[width = 0.33\textwidth]{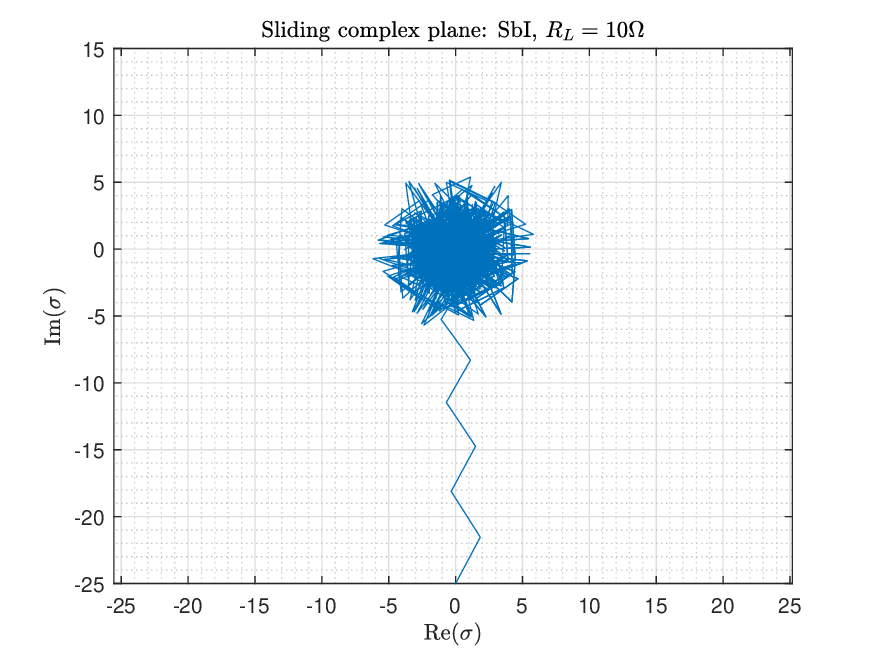}}
\subfloat[Complex Sliding Averaging]{\includegraphics[width = 0.33\textwidth]{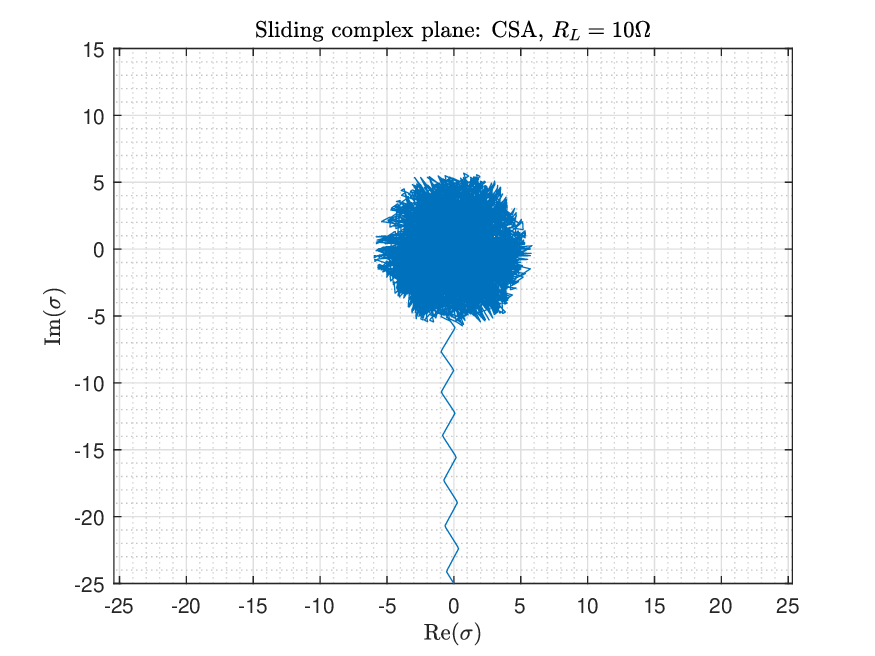}}
\subfloat[Zero-vector Complex Sliding Averaging]{\includegraphics[width = 0.33\textwidth]{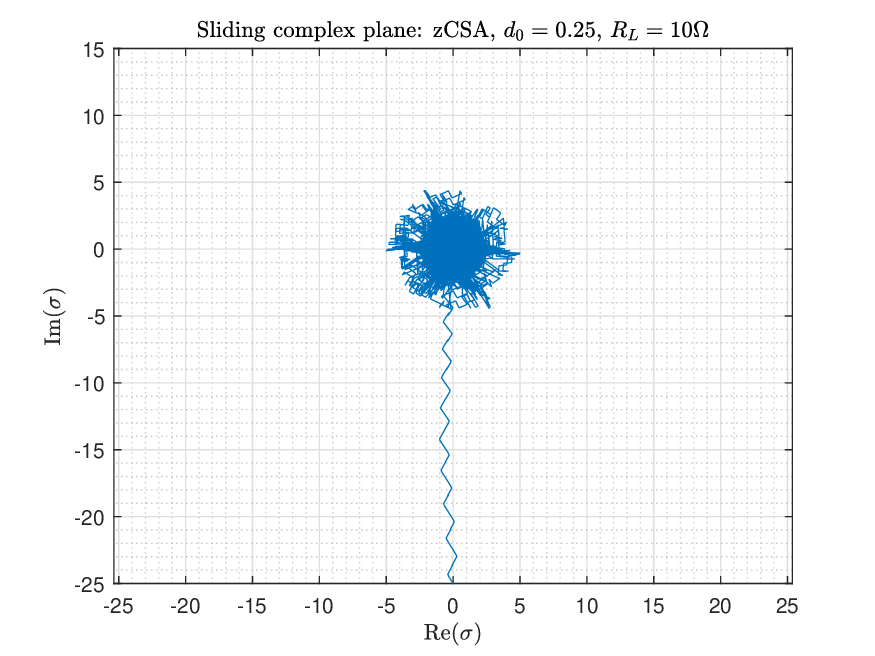}}
\caption{Simulation results: Complex sliding plane for the different implementation methods.}
\label{fig_sims_sig}
\end{figure*}

\begin{figure*}
\centering
\captionsetup[subfloat]{labelfont=scriptsize,textfont=scriptsize}    
\subfloat[Sector-based Implementation]{\includegraphics[width = 0.33\textwidth]{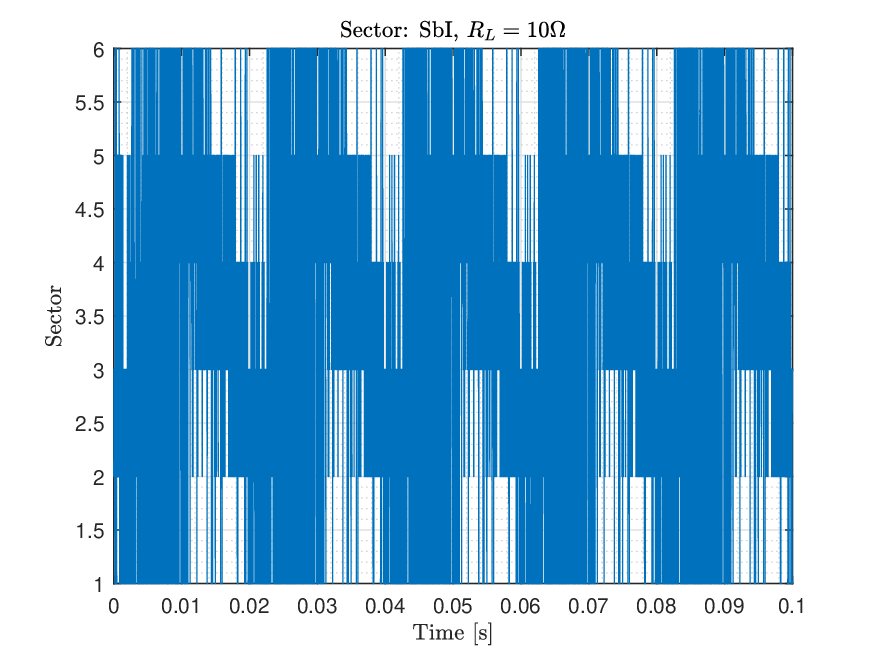}}
\subfloat[Complex Sliding Averaging]{\includegraphics[width = 0.33\textwidth]{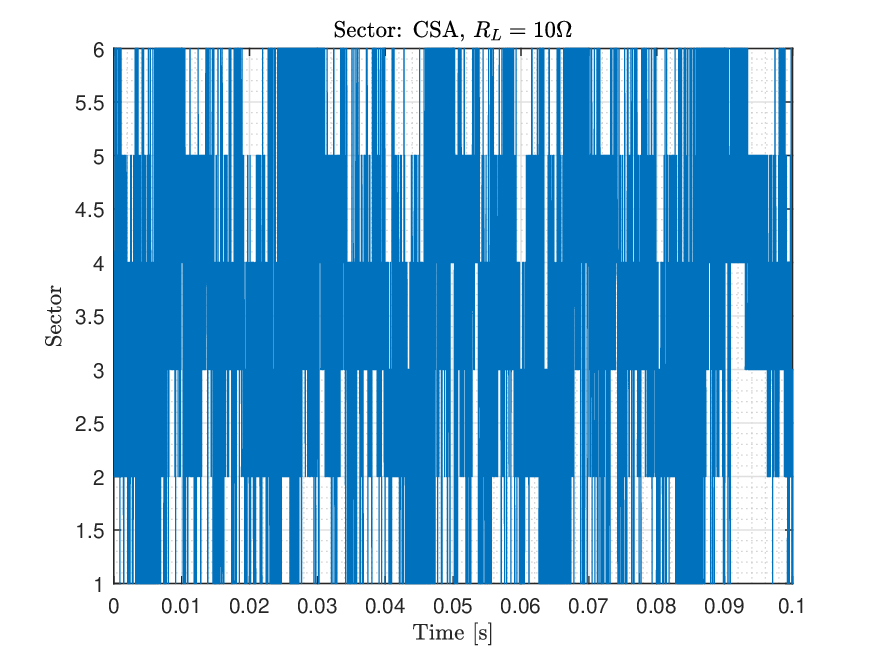}}
\subfloat[Zero-vector Complex Sliding Averaging]{\includegraphics[width = 0.33\textwidth]{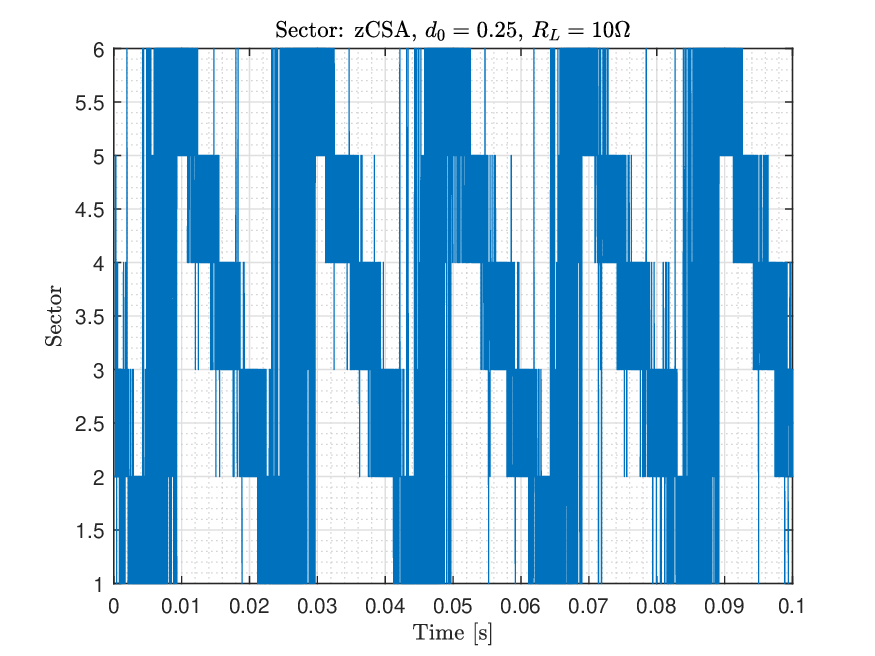}}
\caption{Simulation results: Sectors for the different implementation methods.}
\label{fig_sims_sec}
\end{figure*}

Figures from \ref{fig_sims_transient} to \ref{fig_sims_sec} compare the three implementation methods (SbI, CSA, and zCSA). As shown in Figure \ref{fig_sims_transient}, the three methods track the current references and perform with a fast transient response in front of load and reference changes.

Figure \ref{fig_sims_i} shows the zoom of the three-phase currents at steady-state for the maximum power conditions ($I^\mathrm{ref}=25~\mathrm{A}$ and $R_L=10~\Omega$). Differences are observed in the current ripple; the low ripple current is due to the effective switching frequency of each method. Figure \ref{fig_sims_fft} presents the FFT analysis of the switching control action and shows that zCSA performs in a quasi-fixed frequency given by the sampling frequency, but SbI and CSA have a variable switching frequency. Interestingly, the CSA exhibits frequency peaks at one-third, half, and two-thirds of the sampling frequency. 

Figure \ref{fig_sims_sig} shows the complex sliding plane. Three methods ensure that the trajectories are close to the origin. As expected, the trajectories remain closer to the origin with the zCSA strategy because of the effective gain reduction thanks to zero vectors. Note that the CSA forces $\bm{\sigma}$ to belong to a hexagon. The use of geometric figures has been proposed to implement hysteresis-based methods in complex-values sliding modes \cite{DOBF2021}, which suggests that an alternative to the sampled-based methods proposed in this paper is the use of a complex hysteresis to define the switching events.

Finally, Figure \ref{fig_sims_sec} shows the sector where the desired control action $\bm{u}$ belongs. The zCSA method results in an almost regular pattern, switching between two adjacent sectors, but the sector transitions in the SbI and CSA methods are tangled. 

To evaluate the implementation strategies, the root-mean-square error (RMSE) and the maximum absolute error (MAE) of the currents in one signal period ($20~\mathrm{ms}$) are compared. These Key Performance Indexes (KPIs) are shown in Table \ref{table_KPI} for each method and different zero duty cycles lower than their maximum. The methods CSA and zCSA perform better than SbI, with the best results when the zero duty cycle is $0.3$, close to the maximum, $d_0^\mathrm{max}=0.325$. As expected, for higher values, the sliding motion is lost.


\begin{table}[htbp]
\centering
\begin{tabular}{||c||c|c||}
	\hline 
Method & RMSE [A] & MAE [A]\\ 
	\hline 
SbI & $1.6711$ & $5.8084$ \\ 
	\hline 
CSA & $1.6418$ & $5.7734$ \\ 
	\hline 
zCSA ($d_0=0.05$) & $1.6518$ & $5.6636$ \\ 
	\hline 
zCSA ($d_0=0.10$) & $1.3426$ & $5.3995$ \\ 
	\hline 
zCSA ($d_0=0.15$) & $1.3208$ & $5.2961$ \\ 
	\hline 
zCSA ($d_0=0.20$) & $1.1431$ & $4.3982$ \\ 
	\hline 
zCSA ($d_0=0.25$) & $1.0426$ & $4.2627$ \\ 
	\hline 	
zCSA ($d_0=0.30$) & $\bm{0.9360}$ & \bm{$2.5168$} \\ 
	\hline 	
\end{tabular} 
\caption{Key Performance Indexes.}
\label{table_KPI}
\end{table}

\subsection{Experimental results}\label{sec_Experimental}

\begin{figure}
\centering
\includegraphics[width=8.5cm]{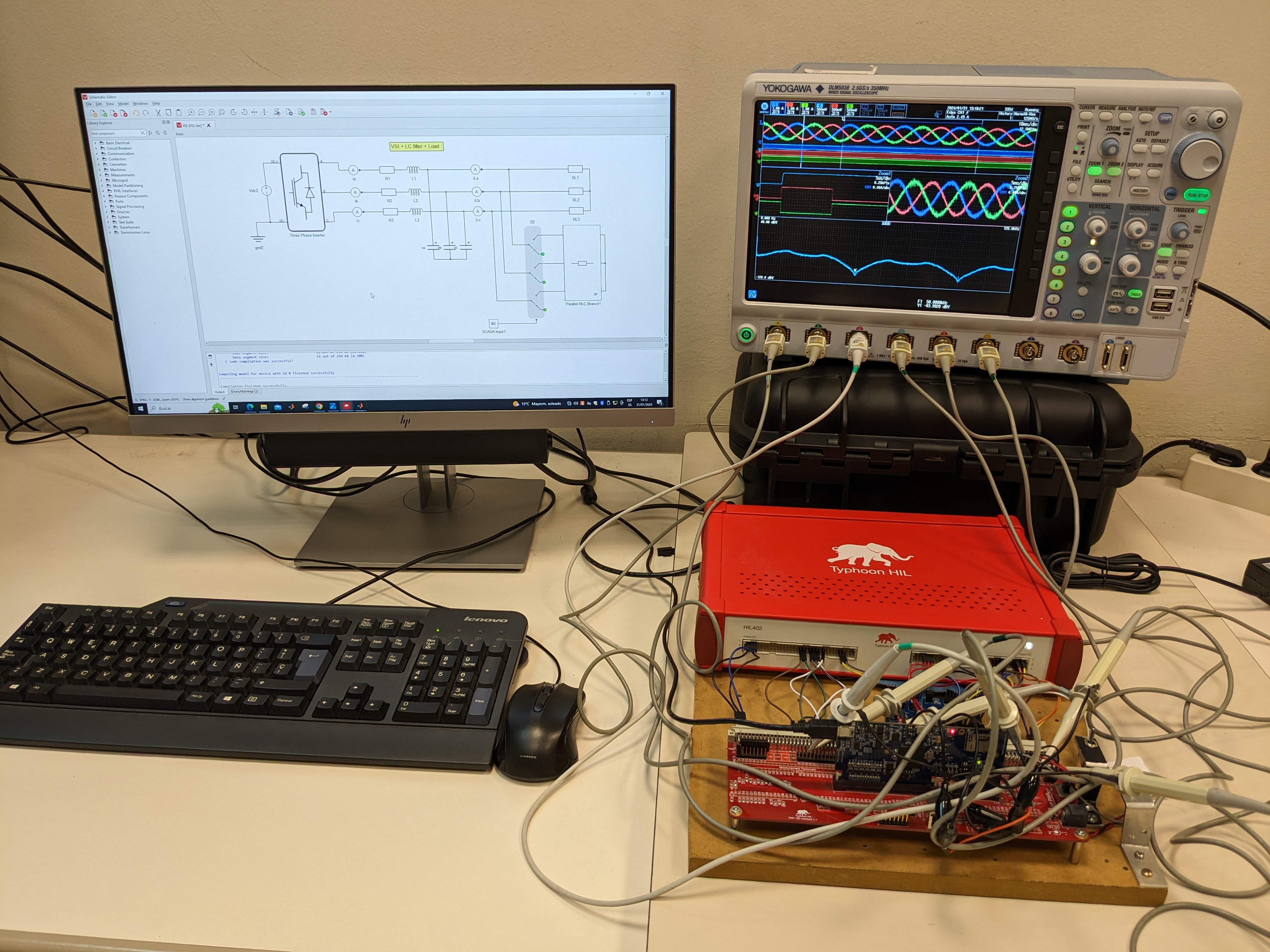}
\caption{Experimental setup: Typhoon HIL 402, HIL DSP 180 interface, and Delfino F28379D from Texas Instruments.}\label{fig_setup}
\end{figure}

The implementation strategies have been experimentally validated in a microcontroller. The setup consists of Typhoon HIL 402, HIL DSP 180 Interface, and Delfino F28379D DSP from Texas Instruments, see Figure \ref{fig_setup}. The VSI parameters, the sampling time, and the current reference were the same as the ones used in the simulation stage.  The PWM peripherals of the microcontroller were used to implement the CSA and zCSA strategies. The programming of the PWM peripheral can be simplified if the order of the vector application is modified such that (along the sampling time) all the rising pulses are centred. According to this, the order of the vectors in the odd sectors (S1, S3 and S5) is $u_{abc}^-,u_{abc}^+,u_{abc}^-$, on the contrary to the one shown in Figure \ref{fig_CSA_carrier}. This modification does not modify the average performance.


The test started with $I^\mathrm{ref}=25~\mathrm{A}$ and $R_L=5~\Omega$ and consisted of a resistance change to $10~\Omega$ and, after approximately $200~\mathrm{ms}$, a reference change to $15~\mathrm{A}$ as the current peak value.

\begin{figure}
\scalebox{1.0}{\centering
\begin{tikzpicture}
    \draw (0, 0) node[inner sep=0] {\includegraphics[width=8.5cm]{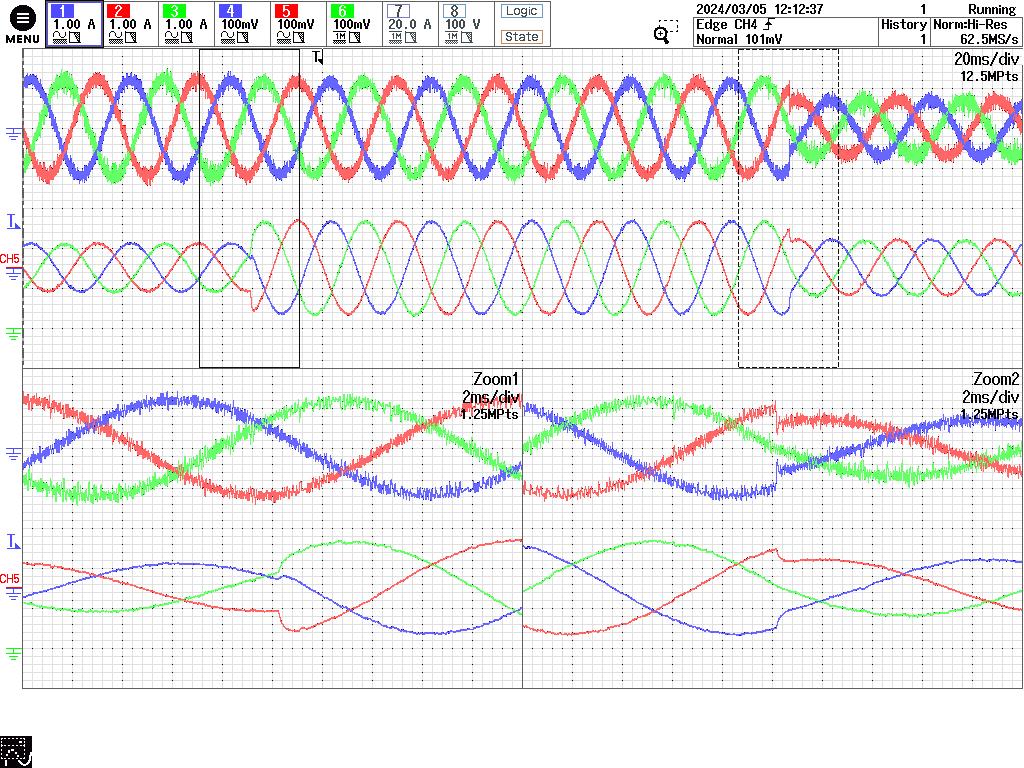}};
    \draw (-4.1, 2.70) node {\tiny $25~\mathrm{A}$};
    \draw (-4.2, 2.55) node {\tiny $\rightarrow$}; 
    \draw (-4.05, 1.30) node {\tiny $100~\mathrm{V}$};
    \draw (-4.2, 1.15) node {\tiny $\rightarrow$};     
\end{tikzpicture}}
\caption{Experimental results for the SbI method: (top) inductor currents, $i_{abc}$ (CH1, CH2, CH3), and capacitor voltages, $v_{abc}$ (CH4, CH5, CH6). (bottom left) Zoom of the load change. (bottom right) Zoom of the reference change.}\label{osc_loadchange_SbI}
\end{figure}

\begin{figure}
\scalebox{1.0}{\centering
\begin{tikzpicture}
    \draw (0, 0) node[inner sep=0] {\includegraphics[width=8.5cm]{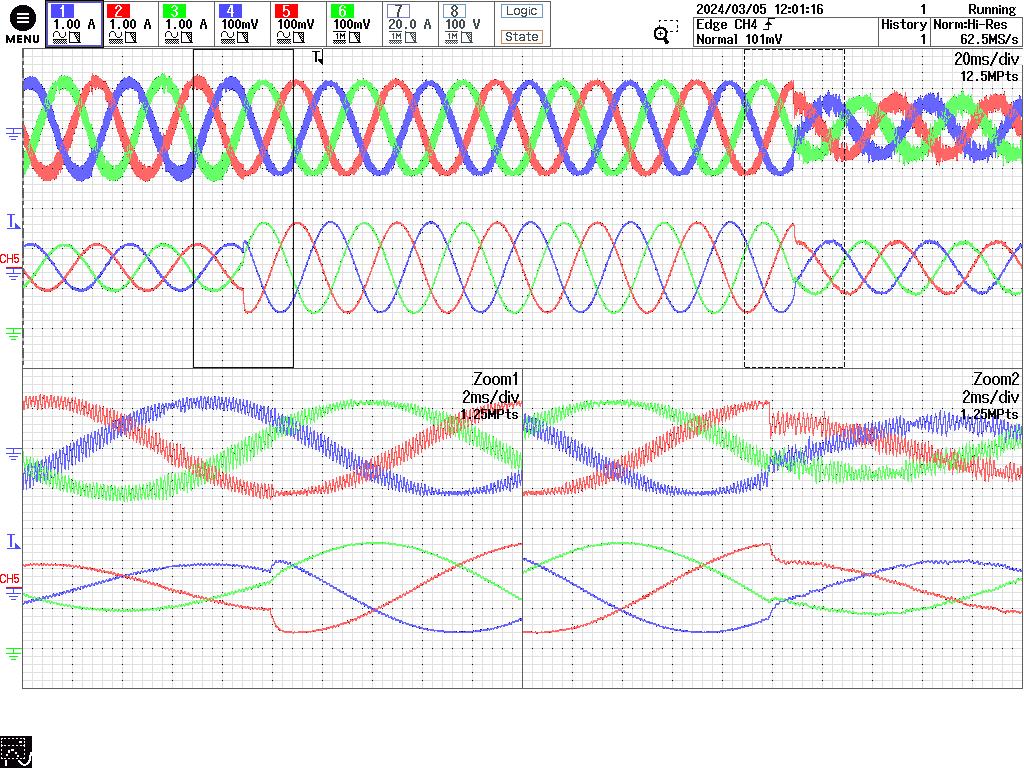}};
    \draw (-4.1, 2.70) node {\tiny $25~\mathrm{A}$};
    \draw (-4.2, 2.55) node {\tiny $\rightarrow$}; 
    \draw (-4.05, 1.30) node {\tiny $100~\mathrm{V}$};
    \draw (-4.2, 1.15) node {\tiny $\rightarrow$};     
\end{tikzpicture}}
\caption{Experimental results for the CSA method: (top) inductor currents, $i_{abc}$ (CH1, CH2, CH3), and capacitor voltages, $v_{abc}$ (CH4, CH5, CH6). (bottom left) Zoom of the load change. (bottom right) Zoom of the reference change. }\label{osc_loadchange_CSA}
\end{figure}

\begin{figure}
\scalebox{1.0}{\centering
\begin{tikzpicture}
    \draw (0, 0) node[inner sep=0] {\includegraphics[width=8.5cm]{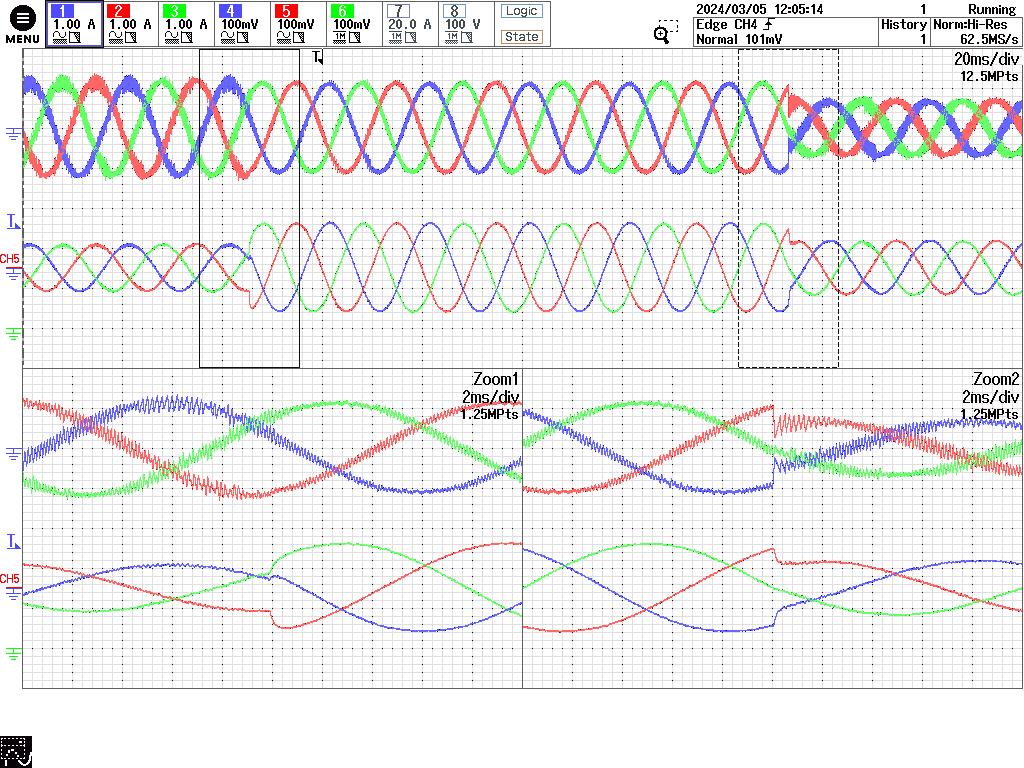}};
    \draw (-4.1, 2.70) node {\tiny $25~\mathrm{A}$};
    \draw (-4.2, 2.55) node {\tiny $\rightarrow$}; 
    \draw (-4.05, 1.30) node {\tiny $100~\mathrm{V}$};
    \draw (-4.2, 1.15) node {\tiny $\rightarrow$};     
\end{tikzpicture}}
\caption{Experimental results for the zCSA method with $d_0=0.25$: (top) inductor currents, $i_{abc}$ (CH1, CH2, CH3), and capacitor voltages, $v_{abc}$ (CH4, CH5, CH6). (bottom left) Zoom of the load change. (bottom right) Zoom of the reference change. }\label{osc_loadchange_zCSA_d025}
\end{figure}

\begin{figure}
\centering\scalebox{1.0}{
\begin{tikzpicture}
    \draw (0, 0) node[inner sep=0] {\includegraphics[width=8.5cm]{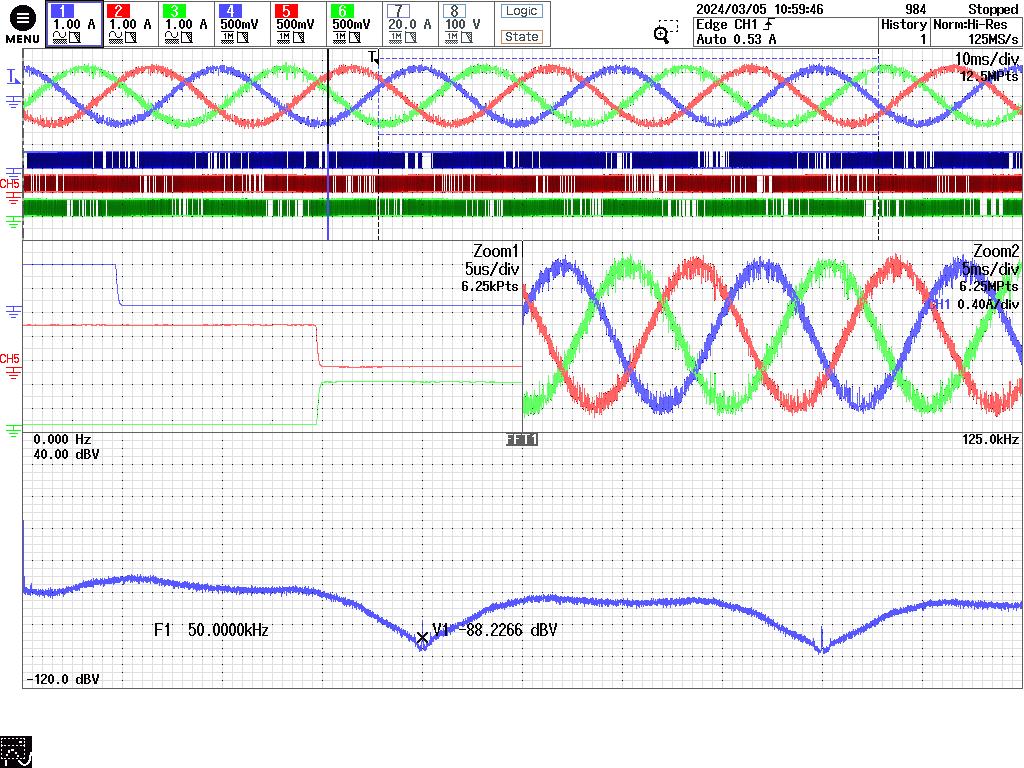}};
    \draw[dashed] (-3.29,-0.35) -- (-3.29,1.15); 
    \draw[dashed] (-1.63,-0.35) -- (-1.63,1.15);    
    \draw[->] (-4.05,-0.35) -- node[above,pos=0.5,rotate=0,text=black,font=\tiny]{S2} (-3.29,-0.35);  
    \draw[<->] (-3.3,-0.35) -- node[above,pos=0.5,rotate=0,text=black,font=\tiny]{S3} (-1.64,-0.35);      
    \draw[<->] (-1.64,-0.35) -- node[above,pos=0.5,rotate=0,text=black,font=\tiny]{S5} (0.1,-0.35);            
\end{tikzpicture}}
\caption{Experimental results for the SbI method: (top) three-phase inductor currents, $i_{abc}$ (CH1, CH2, CH3), and switching control action, $u_{abc}$ (CH4, CH5, CH6). (middle left) Zoom of the switching control actions, $u_{abc}$. (middle right) Zoom of the three-phase inductor currents, $i_{abc}$. (bottom) FFT of $u_a$ (CH4).}\label{osc_iabc_sbi}
\end{figure}

\begin{figure}
\centering\scalebox{1.0}{
\begin{tikzpicture}
    \draw (0, 0) node[inner sep=0] {\includegraphics[width=8.5cm]{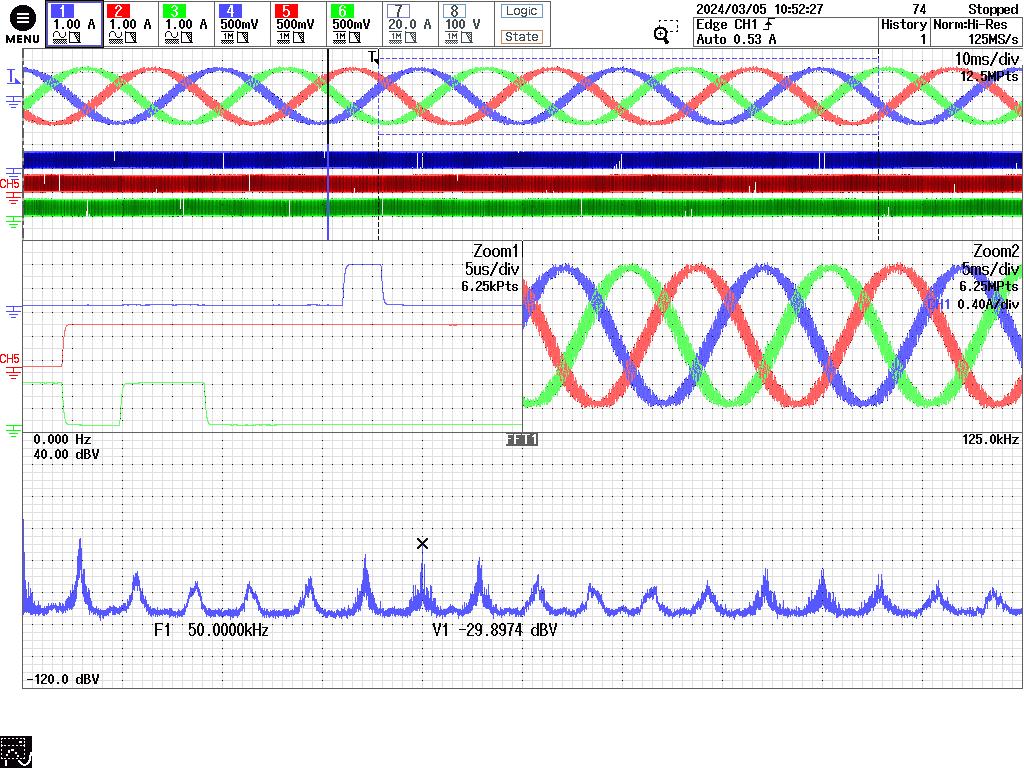}};
    \draw[dashed] (-3.74,-0.35) -- (-3.74,1.15); 
    \draw[dashed] (-2.08,-0.35) -- (-2.08,1.15);    
    \draw[dashed] (-0.42,-0.35) -- (-0.42,1.15);        
    \draw[->] (-4.05,-0.35) -- node[above,pos=0.5,rotate=0,text=black,font=\tiny]{} (-3.74,-0.35);  
    \draw[<->] (-3.74,-0.35) -- node[above,pos=0.5,rotate=0,text=black,font=\tiny]{S3} (-2.08,-0.35);      
    \draw[<->] (-2.08,-0.35) -- node[above,pos=0.5,rotate=0,text=black,font=\tiny]{S2} (-0.42,-0.35);             
    \draw[<-] (-0.42,-0.35) -- node[above,pos=0.5,rotate=0,text=black,font=\tiny]{~} (0.1,-0.35);             
\end{tikzpicture}}
\caption{Experimental results for the CSA method: (top) three-phase inductor currents, $i_{abc}$ (CH1, CH2, CH3), and switching control action, $u_{abc}$ (CH4, CH5, CH6). (middle left) Zoom of the switching control actions, $u_{abc}$. (middle right) Zoom of the three-phase inductor currents, $i_{abc}$. (bottom) FFT of $u_a$ (CH4).}\label{osc_freq_CSA}
\end{figure}

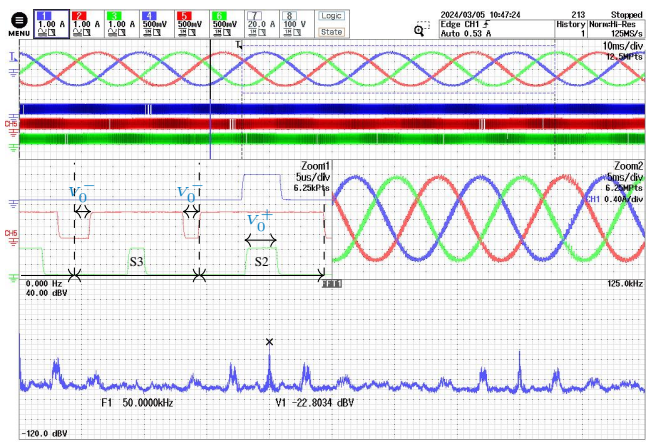
\begin{figure}
\centering\scalebox{1.0}{
\begin{tikzpicture}
    \draw (0, 0) node[inner sep=0] {\includegraphics[width=8.5cm]{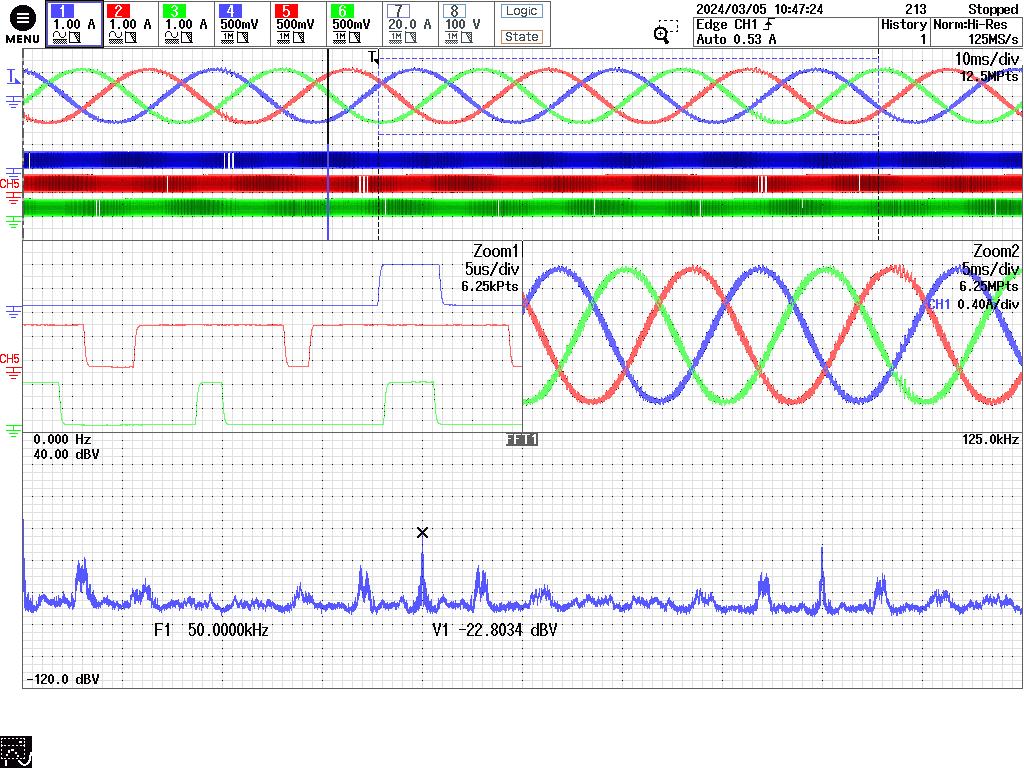}};
    \draw[dashed] (-3.34,-0.35) -- (-3.34,1.15);  
    \draw[dashed] (-1.68,-0.35) -- (-1.68,1.15);  
    \draw[dashed] (-0.02,-0.35) -- (-0.02,1.15);                
    \draw[<->] (-3.34,0.5) -- node[above,pos=0.5,rotate=0,text=UPC,font=\tiny]{$V_0^-$} (-3.12,0.5);
    \draw[<->] (-1.90,0.5) -- node[above,pos=0.5,rotate=0,text=UPC,font=\tiny]{$V_0^-$} (-1.68,0.5);        
    \draw[<->] (-1.08,0.12) -- node[above,pos=0.5,rotate=0,text=UPC,font=\tiny]{$V_0^+$} (-0.65,0.12);       
    \draw[->] (-4.05,-0.35) -- node[above,pos=0.5,rotate=0,text=black,font=\tiny]{~} (-3.34,-0.35);      
    \draw[<->] (-3.34,-0.35) -- node[above,pos=0.5,rotate=0,text=black,font=\tiny]{S3} (-1.68,-0.35);      
    \draw[<->] (-1.68,-0.35) -- node[above,pos=0.5,rotate=0,text=black,font=\tiny]{S2} (-0.02,-0.35);                  
\end{tikzpicture}}
\caption{Experimental results for the zCSA method with $d_0=0.25$: (top) three-phase inductor currents, $i_{abc}$ (CH1, CH2, CH3), and switching control action, $u_{abc}$ (CH4, CH5, CH6). (middle left) Zoom of the switching control actions, $u_{abc}$. (middle right) Zoom of the three-phase inductor currents, $i_{abc}$. (bottom) FFT of $u_a$ (CH4).}\label{osc_freq_zCSA_d025}
\end{figure}

Figures from \ref{osc_loadchange_SbI} to \ref{osc_loadchange_zCSA_d025} show the three-phase inductor currents and the resistor voltages.  Similarly to the simulation results, all the implementations exhibit a fast response when facing the changes, and the main difference is the current ripple. It is worth noticing that the current ripple is minimised when using the zCSA.

Figures from \ref{osc_iabc_sbi} to \ref{osc_freq_zCSA_d025} show the inductor currents, the control signals, and the FFT of the switching control signal for one phase and confirm the results obtained in the simulation stage. The current ripple is reduced when using the CSA and zCSA implementations because the control signals switch closer to the sampling (carrier) frequency effectively. Better results are obtained when the zero duty cycle, $d_0$, is set according to the load/reference conditions. Finally, note how the vectors appear when the zCSA method is used; see the zoom in the middle left in Fig. \ref{osc_freq_zCSA_d025}


\section{Conclusions}\label{sec_concl}

Three strategies for implementing complex-valued sliding mode controllers in three-phase inverters have been proposed. The algorithms are based on approximating the required switching control action by the available ones in the three-phase inverter and work with fixed sampling time. The methods have the following properties:
\begin{itemize}
\item Even though the approximation of the control action, the sliding motion is preserved.
\item The methods are easily programable in the microcontroller and have a low computational burden.
\item The algorithms work with fixed sampling time but result in variable switching frequencies (SbI and CSA) and a quasi-fixed switching frequency (zCSA).
\end{itemize}

Additionally, the CSA and zCSA methods can be applied for converters with different numbers of legs and levels, where the application of sliding control methods is usually difficult.

Future work will study the adaptation of zero-duty time to the working point.


%
%
%


\balance

\bibliographystyle{IEEEtran}
\bibliography{complex_smc-ref}

\begin{thebibliography}{10}
\providecommand{\url}[1]{#1}
\csname url@samestyle\endcsname
\providecommand{\newblock}{\relax}
\providecommand{\bibinfo}[2]{#2}
\providecommand{\BIBentrySTDinterwordspacing}{\spaceskip=0pt\relax}
\providecommand{\BIBentryALTinterwordstretchfactor}{4}
\providecommand{\BIBentryALTinterwordspacing}{\spaceskip=\fontdimen2\font plus
\BIBentryALTinterwordstretchfactor\fontdimen3\font minus
  \fontdimen4\font\relax}
\providecommand{\BIBforeignlanguage}[2]{{%
\expandafter\ifx\csname l@#1\endcsname\relax
\typeout{** WARNING: IEEEtran.bst: No hyphenation pattern has been}%
\typeout{** loaded for the language `#1'. Using the pattern for}%
\typeout{** the default language instead.}%
\else
\language=\csname l@#1\endcsname
\fi
#2}}
\providecommand{\BIBdecl}{\relax}
\BIBdecl

\bibitem{HL2003}
D.~G. Holmes and T.~A. Lipo, \emph{Pulse Width Modulation for Power Converters:
  Principles and Practice}.\hskip 1em plus 0.5em minus 0.4em\relax Wiley-IEEE
  Press, 2003.

\bibitem{UGS1999}
V.~Utkin, J.~Guldner, and J.~Shi, \emph{Sliding Mode Control in
  Electro-Mechanical Systems}.\hskip 1em plus 0.5em minus 0.4em\relax CRC
  Press, 1999.

\bibitem{DORB2020}
A.~D{\`o}ria-Cerezo, J.~M. Olm, V.~Repecho, and D.~Biel, ``Complex-valued
  sliding mode control of an induction motor,'' \emph{IFAC-PapersOnLine},
  vol.~53, no.~2, pp. 5473--5478, 2020.

\bibitem{DHB2022}
A.~D\`{o}ria-Cerezo, M.~A. Hossain, and M.~Bodson, ``Complex-valued sliding
  mode controllers for doubly-fed induction motors,'' \emph{IEEE Trans. on
  Control Systems Technology}, vol.~31, no.~3, pp. 1336--1344, 2022.

\bibitem{SMTSJB2023}
A.~Susperregui, M.~I. Mart{\'\i}nez, G.~Tapia-Otaegui, J.~A. Solsona, S.~G.
  Jorge, and C.~A. Busada, ``Complex-valued sliding-mode control for {DFIG}
  synchronization to non-ideal grids,'' \emph{IFAC-PapersOnLine}, vol.~56,
  no.~2, pp. 2746--2752, 2023.

\bibitem{DRB2021}
A.~D{\`o}ria-Cerezo, V.~Repecho, and D.~Biel, ``Three-phase phase-locked loop
  algorithms based on sliding modes,'' \emph{IEEE Trans. on Power Electronics},
  vol.~36, no.~9, pp. 10\,842--10\,851, 2021.

\bibitem{DOBF2021}
A.~D\`{o}ria-Cerezo, J.~M. Olm, D.~Biel, and E.~Fossas, ``Sliding modes of
  complex-valued nonlinear systems,'' \emph{IEEE Trans. on Automatic Control},
  vol.~66, no.~7, pp. 3355--3362, 2021.

\bibitem{Harnefors2007}
L.~Harnefors, ``Modeling of three-phase dynamic systems using complex transfer
  functions and transfer matrices,'' \emph{IEEE Trans. on Industrial
  Electronics}, vol.~54, no.~4, pp. 2239--2248, 2007.

\bibitem{HDM2013}
D.~G. Holmes, R.~Davoodnezhad, and B.~P. Mcgrath, ``An improved three-phase
  variable-band hysteresis current regulator,'' \emph{IEEE Trans. on Power
  Electronics}, vol.~28, no.~1, pp. 441--450, 2013.

\bibitem{HCC2009}
C.-M. Ho, V.~Cheung, and H.-H. Chung, ``Constant-frequency hysteresis current
  control of grid-connected {VSI} without bandwidth control,'' \emph{IEEE
  Trans. on Power Electronics}, vol.~24, no.~11, pp. 2484--2495, 2009.

\bibitem{MMT1997}
L.~Malesani, P.~Mattavelli, and P.~Tomasin, ``Improved constant-frequency
  hysteresis current control of vsi inverters with simple feedforward bandwidth
  prediction,'' \emph{IEEE Trans. on Industry Applications}, vol.~33, no.~5,
  pp. 1194--1202, 1997.

\bibitem{RBFG2003}
R.~Ramos, D.~Biel, E.~Fossas, and F.~Guinjoan, ``A fixed-frequency
  quasi-sliding control algorithm: Application to power inverters design by
  means of fpga implementation,'' \emph{IEEE Trans. on Power Electronics},
  vol.~18, no.~1, pp. 344--355, 2003.

\bibitem{QLTH2018}
W.~Qi, S.~Li, S.-C. Tan, and S.~Y.~R. Hui, ``Parabolic-modulated sliding-mode
  voltage control of a buck converter,'' \emph{IEEE Trans. on Industrial
  Electronics}, vol.~65, no.~1, pp. 844--854, 2018.

\bibitem{AAM2012}
A.~Abrishamifar, A.~Ahmad, and M.~Mohamadian, ``Fixed switching frequency
  sliding mode control for single-phase unipolar inverters,'' \emph{IEEE Trans.
  on Power Electronics}, vol.~27, no.~5, pp. 2507--2514, 2012.

\bibitem{HYLHC2013}
X.~Hao, X.~Yang, T.~Liu, L.~Huang, and W.~Chen, ``A sliding-mode controller
  with multiresonant sliding surface for single-phase grid connected {VSI} with
  an {LCL} filter,'' \emph{IEEE Trans. on Power Electronics}, vol.~28, no.~5,
  pp. 2259--2268, 2013.

\bibitem{ZJSGT2018}
L.~Zheng, F.~Jiang, J.~Song, Y.~Gao, and M.~Tian, ``A discrete-time repetitive
  sliding mode control for voltage source inverters,'' \emph{IEEE Journal of
  Emerging and Selected Topics in Power Electronics}, vol.~6, no.~3, pp.
  1533--1566, 2018.

\bibitem{VMMS2018}
R.~P. Vieira, L.~T. Martins, J.~R. Massing, and M.~Stefanello, ``Sliding mode
  controller in a multiloop framework for a grid-connected {VSI} with {LCL}
  filter,'' \emph{IEEE Trans. on Industrial Electronics}, vol.~65, no.~6, pp.
  4714--4723, 2018.

\bibitem{PR2017}
M.~Pichan and H.~Rastegar, ``Sliding-mode control of four-leg inverter with
  fixed switching frequency for uninterruptible power supply applications,''
  \emph{IEEE Trans. on Industrial Electronics}, vol.~64, no.~8, pp. 6805--6814,
  2017.

\bibitem{SVKMA2016}
F.~Sebaaly, H.~Vahedi, H.~Y. K.~N. Moubayed, and K.~Al-Haddad, ``Sliding mode
  fixed frequency current controller design for grid-connected {NPC}
  inverter,'' \emph{IEEE Journal of Emerging and Selected Topics in Power
  Electronics}, vol.~4, no.~4, pp. 1397--1405, 2016.

\bibitem{TC2002}
T.-L. Tai and J.-S. Chen, ``{UPS} inverter design using discrete-time
  sliding-mode control scheme,'' \emph{IEEE Trans. on Industrial Electronics},
  vol.~49, no.~1, pp. 67--75, 2002.

\bibitem{RBO2021}
V.~Repecho, D.~Biel, and J.~M. Olm, ``A simple switching-frequency-regulated
  sliding-mode controller for a {VSI} with a full digital implementation,''
  \emph{IEEE Journal of Emerging and Selected Topics in Power Electronics},
  vol.~9, no.~1, pp. 569--579, 2021.

\bibitem{LSGHG2020}
J.~Lu, M.~Savaghebi, A.~M. Y.~M. Ghias, X.~Hou, and J.~M. Guerrero, ``A
  reduced-order generalized proportional integral observer-based resonant
  super-twisting sliding mode control for grid-connected power converters,''
  \emph{IEEE Trans. on Industrial Electronics}, vol.~68, no.~7, pp. 5897--5908,
  2020.

\bibitem{OBBBS2021}
S.~Ouchen, M.~Benbouzid, F.~Blaabjerg, A.~Betka, and H.~Steinhart, ``Direct
  power control of shunt active power filter using space vector modulation
  based on supertwisting sliding mode control,'' \emph{IEEE Journal of Emerging
  and Selected Topics in Power Electronics}, vol.~9, no.~3, pp. 3243--3253,
  2021.

\bibitem{GGMCM2016}
R.~Guzman, L.~G. de~Vicu{\~n}a, J.~Morales, M.~Castilla, and J.~Matas,
  ``Sliding-mode control for a three-phase unity power factor rectifier
  operating at fixed switching frequency,'' \emph{IEEE Trans. on Power
  Electronics}, vol.~31, no.~1, pp. 758--769, 2016.

\bibitem{AOKS2019}
N.~Altin, S.~Ozdemir, H.~Komurcugil, and I.~Sefa, ``Sliding-mode control in
  natural frame with reduced number of sensors for three-phase grid-tied
  {LCL}-interfaced inverters,'' \emph{IEEE Trans. on Industrial Electronics},
  vol.~66, no.~4, pp. 2903--2913, 2019.

\bibitem{RBA2018}
V.~Repecho, D.~Biel, and A.~Arias, ``Fixed switching period discrete-time
  sliding mode current control of a pmsm,'' \emph{IEEE Trans. on Industrial
  Electronics}, vol.~65, no.~3, pp. 2039--2048, 2018.

\bibitem{MGGCM2018}
J.~Morales, L.~{Garcia de Vicu\~na}, R.~Guzman, M.~Castilla, and J.~Miret,
  ``Modeling and sliding mode control for three-phase active power filters
  using the vector operation technique,'' \emph{IEEE Trans. on Industrial
  Electronics}, vol.~65, no.~9, pp. 6828--6838, 2018.

\end{thebibliography}

%

\end{document}